\documentclass[10pt,journal,final]{IEEEtran}
\IEEEoverridecommandlockouts

\usepackage{amsfonts,amssymb}
\usepackage{array}
\usepackage{ifpdf}
\usepackage{cite}
\usepackage{stfloats}
\usepackage{bm}
\usepackage{physics}
\usepackage{verbatim}
\usepackage{color,xcolor}
\usepackage{subfigure}
\usepackage{cite}
\usepackage{subfloat}

\usepackage{lipsum,amsmath,multicol}

\ifCLASSINFOpdf
\usepackage[pdftex]{graphicx}
\graphicspath{{../pdf/}{../jpeg/}}
\DeclareGraphicsExtensions{.pdf,.jpeg,.png}
\else
\usepackage[dvips]{graphicx}

\graphicspath{{../eps/}}

\DeclareGraphicsExtensions{.eps}
\fi

\usepackage{amsmath, amssymb}

\usepackage{algorithm}
\usepackage{algorithmic}

\usepackage{makecell}

\allowdisplaybreaks[4]

\begin{document}
	
	\title{Joint Beamforming Design and Resource Allocation
		for IRS-Assisted Full-Duplex Terahertz Systems}
	\author{Chi~Qiu,~Wen~Chen,~Qingqing~Wu,~Fen~Hou,~Wanming~Hao,~Ruiqi~Liu,~Derrick~Wing~Kwan~Ng\thanks{
	
 C. Qiu is with the Department of Electronic Engineering, Shanghai Jiao Tong University, Minhang 200240, China, and also with   the School of Computer Science and Information Engineering, Hubei University, Wuhan 430062, China (e-mail: chiqiu@hubu.edu.cn). W. Chen and Q. Wu are with the Department of Electronic Engineering, Shanghai Jiao Tong University, Minhang 200240, China (e-mail: wenchen@sjtu.edu.cn; qingqingwu@sjtu.edu.cn). F. Hou is with the State Key Laboratory of Internet of Things for Smart City and the Department of Electrical and Computer Engineering, University of Macau, Macau SAR, 999078, China (e-mail: fenhou@um.edu.mo). W. Hao is with the School of Electrical and Information
  Engineering, Zhengzhou University, Zhengzhou 450001, China (e-mail: iewmhao@zzu.edu.cn). R. Liu is with the Wireless and
  Computing Research Institute, ZTE Corporation, Beijing 100029, China (email: richie.leo@zte.com.cn). D. W. K. Ng is with the School of Electrical Engineering
  and Telecommunications, University of New South Wales, Sydney, Australia
  (e-mail: w.k.ng@unsw.edu.au). Part of this paper was accepted by IEEE ICC Workshops 2025.}}
	\maketitle

	\begin{abstract}
		Intelligent reflecting surface (IRS)-assisted full-duplex (FD) terahertz (THz) communication systems have emerged as a promising paradigm to satisfy the escalating demand for ultra-high data rates and spectral efficiency in future wireless networks. However, the practical deployment of such systems presents unique technical challenges, stemming from severe propagation loss, frequency-dependent molecular absorption in the THz band, and the presence of strong residual self-interference (SI) inherent to FD communications. To tackle these issues, this paper proposes a joint resource allocation framework that aims to maximize the weighted minimum rate  among all users, thereby ensuring fairness in quality of service. Specifically, the proposed design jointly optimizes IRS reflecting phase shifts, uplink/downlink transmit power control, sub-band bandwidth allocation, and sub-band assignment, explicitly capturing the unique propagation characteristics of THz channels and the impact of residual SI. To strike an balance between system performance and computational complexity, two computationally  efficient algorithms are developed under distinct spectrum partitioning schemes: one assumes equal sub-band bandwidth allocation to facilliate tractable optimization, while the other introduces adaptive bandwidth allocation to further enhance spectral utilization and system flexibility.  Simulation results validate the effectiveness of the proposed designs and demonstrate that the adopted scheme achieves significant  spectral efficiency improvements over benchmark schemes.
	\end{abstract}


	\section{Introduction}
The terahertz (THz) band, typically spanning frequencies from 0.1 to 10 THz, is widely recognized as a key enabler for sixth-generation (6G) wireless communication systems due to its ultra-wide unlicensed available bandwidth and capability to support exceptionally high data rates. This unprecedented spectral resource opens the door to cope with emerging bandwidth-hungry applications, such as ultra-high-definition holographic transmission, real-time digital twin interactions, and high-precision integrated sensing and communication \cite{hao2022ultra,ning2021prospective,wu2024intelligent,lrq,chen2019survey}. However, the practical implementation of THz communication faces significant challenges arising from the fundamental physics of wave propagation at such high frequencies. Specifically, the THz band is characterized by extremely high free-space path loss, pronounced frequency-related molecular absorption, and vulnerability to signal blockage due to the quasi-optical propagation nature of THz waves \cite{jornet2011channel,han2016distance,hao2022ultra}. These factors collectively result in limited communication ranges, stringent directionality requirements, and frequency-selective channel conditions.

To address these challenges, intelligent reflecting surfaces (IRSs) have been proposed as a disruptive technology for significantly enhancing the robustness and efficiency of wireless systems operating in the THz regime \cite{wu2021intelligent,hua2023secure,zhao2023intelligent}. Specifically, an IRS is a nearly passive metasurface composed of a large number of individually tunable elements capable of manipulating the phase shifts of incident electromagnetic waves. Through the intelligent control of these phase shifts, an IRS can effectively steer reflected signals towards desired locations, reinforce signal strength via passive beamforming, and mitigate interference by redirecting energy away from unintended receivers. Due to their inherent advantages of low cost, power efficiency, and deployment flexibility, IRSs are particularly attractive for applications in the THz frequency band \cite{hou2025220,ning2021terahertz,su2023wideband}. In practice, IRSs can compensate for coverage holes, circumvent line-of-sight (LoS) obstructions, and improve  channel quality for both uplink (UL) and downlink (DL) transmissions.

In addition to the high path loss and sensitivity to blockage, the THz band exhibits a distinct characteristic of pronounced frequency-dependent molecular absorption, primarily attributed to water vapor and oxygen \cite{abs1,abs2}. This absorption results in non-uniform attenuation across the spectrum, effectively dividing the available THz band into a series of irregularly spaced transmission windows (TWs). In particular, within each window, signals experience relatively low attenuation, while the absorption peaks between windows render certain frequencies virtually unusable for long-range communication. Consequently, the available bandwidth in THz systems often becomes fragmented, and varies significantly with environmental conditions such as humidity and distance. This irregular spectral landscape imposes significant constraints on resource allocation and scheduling, since only specific parts of the spectrum can be effectively exploited for communication at any given time. In such a fragmented spectral environment, efficient utilization of the available TWs becomes critical. Thus, a promising strategy is to enhance spectral efficiency within each usable sub-band. This consideration naturally leads to the exploration of full-duplex (FD) communication, which allows simultaneous UL and DL transmissions over the same frequency band, thereby potentially doubling the spectral utilization within the constrained TWs \cite{hao2024resource,hua2022throughput,qiu2023intelligent}. In the context of THz communication, where spectrum is abundant yet practically constrained due to absorption effects and hardware limitations, FD provides a compelling method for boosting throughput without requiring spectral resources.

Despite its advantages in spectral efficiency, practical FD implementation is hindered by the problem of self-interference (SI), whereby a transceiver’s own transmit signal substantially leaks into its receiver chain, severely degrading reception quality. Despite extensive efforts to suppress SI through active radio-frequency (RF) chain techniques \cite{everett2011empowering,fukui2020analog,sahai2011pushing} or in passive manners \cite{qiu2023intelligent,SI}, residual SI typically persists at levels sufficient to impair system performance. Moreover, this residual SI exhibits frequency-dependent characteristics, especially in the THz band, rendering it crucial to account for its impact when optimizing resource allocation in FD THz communication systems.

To tackle these limitations, existing works have proposed IRS-aided designs for FD systems and separately addressed resource allocation in THz networks. For example, IRS technology was deployed for mitigate SI and enhance transmission quality in FD wireless systems \cite{qiu2023intelligent}. Furthermore,  a novel joint dynamic and resource allocation scheme was
proposed to maximize system throughput of the IRS-assisted
FD wireless-powered communication network (WPCN) \cite{hua2021joint}. While these studies demonstrate the effectiveness of IRSs in improving FD system performance, they primarily focus on conventional microwave or sub-6 GHz frequency bands and therefore fail to adequately capture the unique propagation characteristics of the THz environment. On the other hand, a number of studies have investigated resource allocation strategies tailored to THz communications, particularly to address issues such as severe path loss, molecular absorption, and hardware constraints. For example, a sub-carrier scheduling strategy considering molecular absorption effects was proposed in \cite{subcarrier}. Similarly, the joint active and passive beamforming strategies were proposed for multi-IRS-aided multi-user multiple-input multiple-output (MIMO) systems in \cite{ning2022multi}. In addition, resource allocation strategies in STAR-IRS-assisted THz communications was investigated in \cite{yan2024wideband}. However, these studies typically adopt uniform bandwidth or sub-carrier allocation strategies, which do not fully exploit the unique characteristics of THz channels. In fact, the molecular absorption and non-uniform path losses make different parts of the spectrum experience vastly different propagation conditions. Consequently, these existing designs did not fully explore the potential of THz communication systems, particularly in heterogeneous environments, where spectral windows are irregular and  user requirements are diverse. 

These limitations motivate the adoption of adaptive sub-band bandwidth (ASB) allocation, where the available spectrum of interest is partitioned into sub-bands with flexible and user-specific bandwidth, each tailored to varying propagation conditions and service requirements. For instance, in \cite{shafie2021spectrum}, a resource allocation strategy was proposed for THz communications, jointly optimizing multi-band spectrum allocation with ASB and power allocation. This idea was further extended in \cite{ali2024joint} to IRS-assisted multi-user MIMO THz systems. Although these works highlight the benefits of ASB allocation in effectively exploiting the fragmented THz systems, they primarily considered half-duplex communication and thus did not consider the challenges introduced by FD communications, such as frequency-dependent residual SI.  This frequency-selective interference complicates resource allocation and renders existing algorithms inadequate, necessitating new designs that account for SI variations across sub-bands.

Motivated by these conditions, we investigate the joint beamforming design and resource allocation in an IRS-assisted FD THz communication system, with a primary focus on spectrum allocation. Specifically, we consider an FD base station (BS) communicating concurrently with multiple FD users in the THz band, aided by an IRS.
The spectrum of interest is partitioned into multiple sub-bands, with each sub-band exclusively allocated to one user. Crucially, the bandwidth of each sub-band is adaptively optimized, allowing for an enhanced matching between frequency resources and the unique propagation environment, improving both spectral utilization efficiency and user fairness. Indeed, the joint design of ASB allocation, power allocation, and IRS-aided passive beamforming introduces new challenges, including the non-convex coupling among sub-band selection, power control, and IRS phase shifts, that are addressed in this work. 

In this paper, we formulate a weighted minimum rate (WMR) maximization problem that jointly optimizes four key resource dimensions: IRS phase shifts, UL and DL transmit powers, sub-band bandwidth allocation, and sub-band assignment. The formulated problem explicitly captures the inherent coupling between  the frequency-dependent resource distribution and the detrimental effects of residual SI. To strike a balance between tractability and performance, we consider two resource allocation strategies: one with equal sub-band bandwidth (ESB), which simplifies the optimization by avoiding the integration over frequency-dependent channels, and another with ASB to fully exploit the spectral diversity of the THz band. For each resource allocation strategy, we develop efficient algorithms based on convex optimization theory and successive convex approximation (SCA) techniques to obtain high-quality solutions under practical system constraints. The main contributions of this paper are summarized as follows:\begin{itemize}
	\item We investigate explicitly the resource allocation design for a novel IRS-assisted FD THz system architecture, considering frequency-dependent residual SI, channel variations, and fairness among users. Specifically, we formulate a WMR maximization problem, jointly optimizing IRS reflection phase shifts, UL/DL power allocation, sub-band bandwidth allocation, and sub-band assignments. To manage the computational complexity arising from the frequency-dependent nature of THz channels, where rate expressions involve intractable integrals over frequency, we first develop a tractable design with ESB. This simplifies the optimization problem while still capturing the essential spectral variation of the THz band. Under the assumption of perfect SI cancellation, we propose an efficient algorithm capitalizing on SCA that recasts the problem into a convex form.
	
	\item Building upon this simplified approach, we then develop an extended algorithm that incorporates ASB allocation, enabling more efficient utilization of the fragmented THz spectrum. Specifically, a two-layer penalty-based method is proposed, which includes an inner layer that solves a penalized optimization problem exploiting the block coordinate descent (BCD) technique, while an outer layer dynamically updates the penalty coefficient to gradually enforce the penalized equality constraints until convergence is achieved.
	
	\item Simulation results validate the effectiveness and practicality of the proposed framework under realistic THz channel conditions.  In particular, we show that the proposed resource allocation strategies achieve substantial performance improvements in terms of the WMR, especially when the THz channel exhibits strong frequency-related attenuation due to molecular absorption. Furthermore, the critical role of ASB allocation scheme is highlighted as compared to the ESB baselines, as it allows a more effective bandwidth resources matching  to channel conditions. This adaptive approach leads to improved fairness among users and  optimizes the utilization of the fragmented THz spectrum, thus highlighting the practical importance of ASB allocation in IRS-assisted FD THz systems.
\end{itemize}

The remainder of this paper is organized as follows. Section II introduces the system model and problem formulation. Section III proposes an efficient resource allocation algorithm based on ESB, and Section IV extends the investigation to that with ASB. Section V provides simulation results to validate the superiority of the proposed resource allocation strategies. Finally, Section VI concludes the paper.

\emph{Notations}: Throughout this paper, we denote matrices and vectors by boldface upper-case and lower-case letters, respectively. Let $\mathbb{C}^{ a \times b}$ denote the set of complex-valued matrices with dimensions $a\times b$. For a complex-valued
vector $\bm x$,  $\left[\bm x\right]_n$, $(\bm x)^H$, and $\operatorname{diag}(\bm x)$  represent the $n$-th entry, the Hermitian
transpose, and the diagonal matrix formed by the elements of $\mathbf{x}$, respectively. For a complex scalar $x$, $|x|$ and $\operatorname{Re}\{x\}$ denote its absolute value and the real part, respectively.  $\mathcal{O}(\cdot)$ denotes the big-O computational complexity.
	
	\begin{figure}[t]
	\centering
	\includegraphics[width=3in]{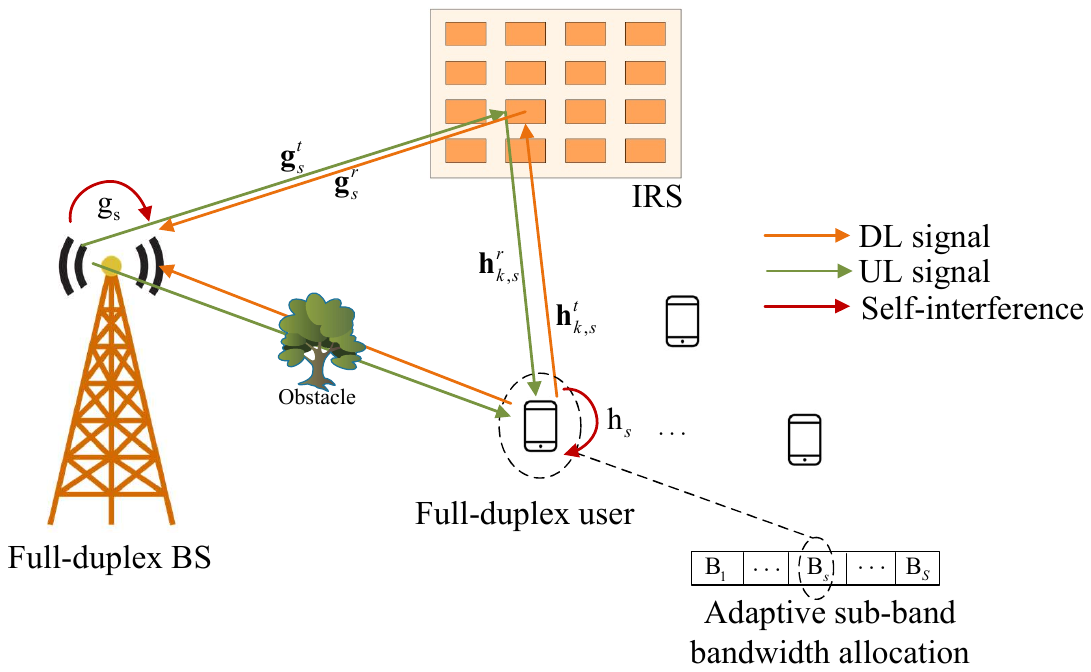}
	\vspace{-0.1cm}
	\caption{An IRS-assisted multi-user FD THz communication system.}
	\label{sys_mod}
	\vspace{-0.3cm}
	\end{figure}

	\section{System Model and Problem Formulation}
	We consider an IRS-assisted multi-user FD THz communication system, as shown in Fig. \ref{sys_mod}, where one BS simultaneously serves $K$ users with the assistance of an IRS. Let $\mathcal{K}=\{1,2,\dots,K\}$ denote the set of users. We assume that both the BS and all users operate in FD mode, where the UL and DL transmissions are concurrently performed over the same frequency band. Following the separate antenna architecture as in \cite{aslani2019energy}, the BS and  users are all equipped with two antennas, where dedicated transmit and receive antennas are exploited for UL and DL signal transmissions, respectively. The IRS contains $N=N_x\times N_y$ reflecting elements, denoted by $\mathcal{N}=\{1,2,\dots,N\}$. Without loss of generality, we assume that the direct communication links between the BS and users are blocked completely by obstacles, which is a common scenario in the IRS-assisted THz communication systems \cite{su2023wideband,subcarrier}. The IRS phase-shift matrix is denoted by\begin{equation}
		\boldsymbol{\Theta}=\operatorname{diag}\left(\mathbf{v}\right)=\operatorname{diag}\left(\left[e^{j\theta},\dots,e^{j\theta_N}\right]\right),
	\end{equation}where $\mathbf{v}$ denotes the IRS passive beamforming vector and $\theta_n,n\in\mathcal{N}$ denotes the phase shift of the $n$-th IRS reflecting element.
		
		\subsection{THz Spectrum}
\begin{figure}[t]
		\centering
		\includegraphics[width=3.2in]{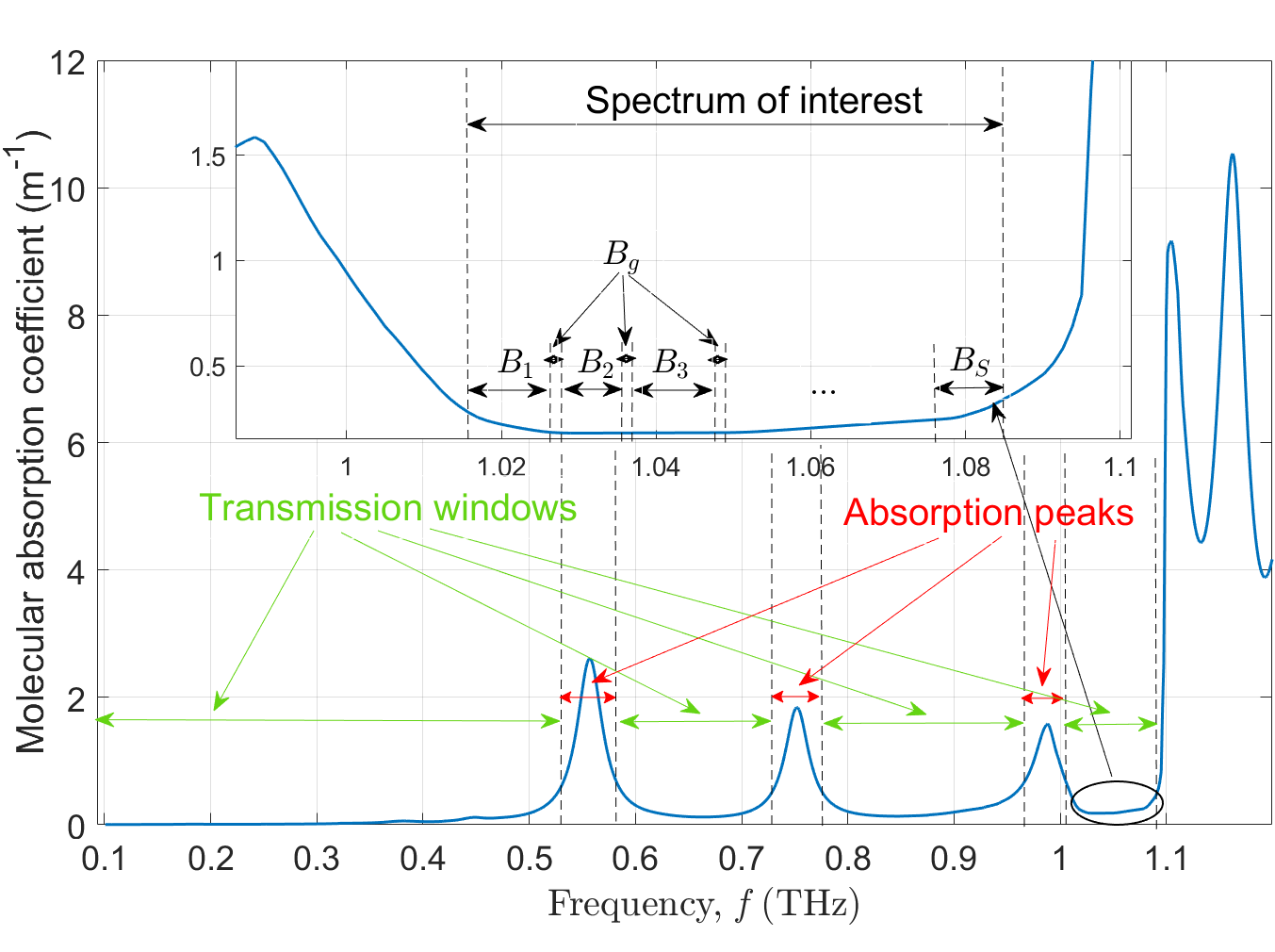}
		\vspace{-0.1cm}
		\caption{Illustration of THz TWs, absorption coefficient peaks, and the allocation of sub-bands.}
		\label{spectrum}
		\vspace{-0.2cm}
	\end{figure}

The ultra-wide bandwidth offered by the THz spectrum enables it to provide extremely high data rate. However, a major challenge is the existence of intermittent molecular absorption peaks \cite{han2016distance,shafie2021spectrum}, which lead to severe path loss and fragmentation of the entire THz band into THz TWs, as shown in Fig. \ref{spectrum}. Since the molecular absorption with each TW is relatively low, as compared to the peaks, these windows can be effectively utilized for signal transmission. Considering this, we focus on the non-overlapping ASB allocation in a TW in this paper. Thus, we consider that the TW of interest is divided into $S$ sub-bands with unequal bandwidths, each of which is separated by a fixed guard band, $B_g$, to avoid inter-band interference \cite{shafie2021spectrum}. Let $\mathcal{S}=\{1,2,\dots,S\}$ denote the set of sub-bands. Furthermore, let  $B=\{B_1,B_2,\dots, B_S\}$ and $f=\{f_1,f_2,\dots,f_S\}$ denote the bandwidths and the central frequencies of the sub-bands, respectively. Then, we have the following constraint for the bandwidth of each sub-band:\begin{equation}\label{B_const0}
		0\leq B_{s}\leq B_{\text{max}}, s\in\mathcal{S},
	\end{equation}where $B_{\text{max}}$ is the maximum allowed bandwidth per sub-band.
	Let $f_{\text{start}}$ and $f_{\text{end}}$ denote the start and end frequency of the TW of interest, respectively. For notational convenience, we assume that the sub-bands are labeled such that $f_1\leq f_2 \leq \dots \leq f_S$. Hence, the central frequency of each sub-band can be expressed as \begin{equation}\label{f_B}
	f_s=f_{\text{start}}+\sum_{i=1}^{s-1}B_{i}+(s-1)B_g+0.5B_{s}.
	\end{equation} Then, the constraint for the total available bandwidth is given by \looseness=-1\begin{equation}\label{Btot_const0}
		 \sum_{s=1}^{S}B_{s}= f_{\text{end}}-f_{\text{start}}-(S-1)B_g.
	\end{equation}
	
	\subsection{Channel Model}
	Signal attenuation is typically high in THz bands and thus scattered components are generally negligible. As a result, we consider the LoS channel model as in \cite{su2023wideband,han2016distance}. Let $\mathbf{g}^{t\,H}\in\mathbb{C}^{1\times N}$ and $\mathbf{h}^{r}_{k}\in\mathbb{C}^{N\times 1}$ denote the DL channels from the BS to the IRS and from the IRS to user $k$, respectively. Considering the free-space spreading and molecular absorption losses, they can be expressed as \cite{ning2021terahertz}\begin{equation}
		\mathbf{g}^{t\,H}\left(f\right)=\frac{c}{4\pi f d_{r}}e^{-\frac{1}{2}\kappa\left(f\right)d_{r}}\mathbf{a}^H\left(f,\psi_r^r,\varphi_r^r\right),
	\end{equation}and\begin{equation}
	\mathbf{h}^{r}_{k}\left(f\right)=\frac{c}{4\pi f d_{k}}e^{-\frac{1}{2}\kappa\left(f\right)d_{k}}\mathbf{a}\left(f,\psi_k^t,\varphi_k^t\right),
	\end{equation}where $d_r$ and $d_{k}$ denote the distances from the BS to IRS, and that from IRS to user $k$, respectively.  $c$ is the speed of light. $\psi_r^r\in \left[-\frac{\pi}{2},\frac{\pi}{2}\right]$ and  $\varphi_r^r\in\left[0,\pi\right]$ are the azimuth and elevation angles of arrival (AoA) of the BS-IRS channel.  $\psi_k^t\in \left[-\frac{\pi}{2},\frac{\pi}{2}\right]$ and $\varphi_k^t\in\left[0,\pi\right]$ are the azimuth and elevation angles of departure (AoD) of the IRS-user $k$ channels. The term $\kappa\left(f\right)$ is the molecular absorption coefficient at frequency $f$, which depends on environmental parameters, such as pressure, temperature, molecular density, and the absorption cross section for different types of isotopologue in different gases. These parameters can be obtained from the HITRAN database \cite{HITRAN}. Thus, the term $e^{-\frac{1}{2}\kappa\left(f\right)d}$ represents the molecular absorption.  Furthermore, $\mathbf{a}\in\mathbb{C}^{N\times 1}$ is the steering vector of the UPA structured IRS for the generic azimuth and elevation angles $\psi$ and $\varphi$, which is expressed as\begin{align}
	\mathbf{a}\left(f,\psi,\varphi\right)=&\left[1,\dots,e^{j2\pi \Delta_d\frac{f}{c} \left(N_x-1\right)\sin(\psi)\sin(\varphi)}\right]\nonumber\\
	&\otimes\left[1,\dots, e^{j2\pi \Delta_d\frac{f}{c} \left(N_y-1\right)\cos(\varphi)}\right],
	\end{align}
	where $\Delta_d$ is the IRS element spacing, $N_x$ denotes the number of IRS elements along the horizontal direction, and $N_y$ denotes that along the vertical direction.
	
	Similarly, let  $\mathbf{h}^{t}_{k}\in\mathbb{C}^{N\times 1}$ and $\mathbf{g}^r\in\mathbb{C}^{N\times 1}$ denote the UL channels from user $k$ to the IRS and from the IRS to the BS, respectively,  which are expressed as\begin{equation}
		\mathbf{h}^{t}_{k}\left(f\right)=\frac{c}{4\pi f d_{k}}e^{-\frac{1}{2}\kappa\left(f\right)d_{k}}\mathbf{a}\left(f,\psi_k^r,\varphi_k^r\right),
	\end{equation}	and\begin{equation}
	\mathbf{g}^{r}\left(f\right)=\frac{c}{4\pi f d_{r}}e^{-\frac{1}{2}\kappa\left(f\right)d_{r}}\mathbf{a}\left(f,\psi_r^t,\varphi_r^t\right),
	\end{equation}where the angles $\psi_k^r$, $\varphi_k^r$,  $\psi_r^t$, and $\varphi_r^t$ are defined in a manner analogous to the previously described DL scenario.
	
	
	In addition, let $g$ and $h_{k}$ denote the  loop SI (LSI) channels of the BS and user $k$, respectively, representing signal leakage from the transmit antenna directly into the receive antenna, which are modeled as\footnote{Perfect SI channel information is assumed available at the BS/users via using pilot-based channel estimation methods \cite{day2012full}.} \begin{equation}
		g\left(f\right)=\sqrt{\beta_{\text{LSI}}}\,e^{-\frac{1}{2}\kappa\left(f\right)d_{k}},
	\end{equation}and\begin{equation}
	h_{k}\left(f\right)=\sqrt{\beta_{\text{LSI}}}\,e^{-\frac{1}{2}\kappa\left(f\right)d_{k}},
	\end{equation}where $\beta_{\text{LSI}}\left(d\right)$ is the path gain of the LSI channel. $d_0$ and $d_{0,k}$ are the antenna spacing of the BS and user $k$. Similar to \cite{shen2020beamformig,hua2021joint}, we neglect the reflected SI via the IRS due to its negligible impact, resulting from the double path loss effect and the passive nature of the IRS elements.  
	It should be noted that in practice, SI cannot be canceled completely even if the SI channel is perfectly known at the BS and  users, due to the limited dynamic range of the receiver. Thus, considering a combination of passive and active SI cancellation methods for all the users and the BS, we assume that the residual SI is proportional to the transmit power \cite{hua2021joint,aslani2019energy}. As a result,  the achieved DL rate of user $k$ in the $s$-th sub-band is given by \cite{shafie2021spectrum,han2016distance} \begin{align}\label{DL_rate}
		R_{k,s}^{\text{d}}=\int_{f_s-\frac{B_s}{2}}^{f_s+\frac{B_s}{2}}\log_2\left(1+\frac{p_{s}\left|\mathbf{h}_{k}^{r\, H}\left(f\right)\boldsymbol{\Theta}\mathbf{g}^{t}\left(f\right)\right|^2}{ \Gamma q_{k,s}\left|h_{k}\left(f\right)\right|^2+N_0}\right)\dd f,
	\end{align} where  $p_{s}$ is the BS transmit power in the $s$-th sub-band, $q_{k,s}$ is the transmit power of user $k$ in the $s$-th sub-band, $0\leq\Gamma\ll 1$ is the quality of SI cancellation, and $N_0$ is the noise power density. 
	
	
	Similarly, the achieved UL rate of user $k$ in the $s$-th sub-band is given by  \begin{align}\label{UL_rate}
		R_{k,s}^{\text{u}}=\int_{f_s-\frac{B_s}{2}}^{f_s+\frac{B_s}{2}}\log_2\left(1+\frac{q_{k,s}\left|\mathbf{g}^{r}\left(f\right)\boldsymbol{\Theta}\mathbf{h}_{k}^t\left(f\right)\right|^2}{ \Gamma p_{s}\left|g\left(f\right)\right|^2+N_0}\right)\dd f.
	\end{align}
	
	\vspace{-0.2cm}
	\subsection{Problem Formulation}
	We assume that each sub-band is allocated to one user exclusively such that multi-user interference and co-channel interference can be avoided. Specifically, we introduce an indicating parameter $\alpha_{k,s}$, such that $\alpha_{k,s}=1$ if sub-band $s$  is allocated to user $k$, and denote $\mathbf{\alpha}=\{\alpha_{k,s}, k\in\mathcal{K},s\in\mathcal{S}\}$.
We aim to maximize the WMR, denoted by $\tau\triangleq \min\limits_{k\in\mathcal{K},i\in\{\text{d,u}\}}\omega_{i,k}\sum\limits_{s\in\mathcal{S}}\alpha_{k,s}R^i_{k,s}$, where $\omega_{i,k}$ is the weighting factor, to guarantee the fairness among users, by jointly optimizing the IRS phase shifts,  DL/UL transmission power allocation, sub-band bandwidth allocation, and sub-band assignment. Denote $p=\{p_{s},s\in\mathcal{S}\}$ and $q=\{q_{k,s},k\in\mathcal{K},s\in\mathcal{S}\}$. The WMR maximization problem is formulated as\begin{subequations}\label{p1}
		\begin{align}
			& \max_{B,\mathbf{v},p,q,\mathbf{\alpha},\tau}\quad   \tau  \\
			&\text{s.t.} \quad \omega_{i,k}\sum_{s=1}^{S}\alpha_{k,s}R^{i}_{k,s}\geq \tau, i\in\{\text{d,u}\}, k\in\mathcal{K},\label{rate_const}\\
			&\quad\quad\alpha_{k,s}=\{0,1\},  k\in\mathcal{K},s\in\mathcal{S},\label{binary}\\
			&\quad\quad\sum_{k=1}^{K}	\alpha_{k,s}\leq1, s\in\mathcal{S}, \label{subband_const}\\
			&\quad\quad\sum_{s=1}^{S}\sum_{k=1}^{K}\alpha_{k,s}p_{s}\leq P_{\text{b}},\label{DL_P_const}\\
			&\quad\quad\sum_{s=1}^{S}\alpha_{k,s}q_{k,s}\leq P_{k}, k\in \mathcal{K}\label{UL_P_const}\\
			&\quad\quad p_{s} \geq 0,q_{k,s}\geq 0, s\in\mathcal{S}, k\in \mathcal{K},\label{pos_P_const}\\
			& \quad\quad0\leq B_{s}\leq B_{\text{max}}, s\in\mathcal{S},\label{B_const}\\
			&\quad\quad\sum_{s=1}^{S}B_{s}= f_{\text{end}}-f_{\text{start}}-(S-1)B_g,\label{Btot_const}\\
			&\quad\quad|[\mathbf{v}]_n|=1, n\in\mathcal{N}.\label{unit_mod_const}
		\end{align}
	\end{subequations}Constraint \eqref{binary} is the integer constraint for sub-band assignment. Constraint \eqref{subband_const} ensures  that each sub-band is allocated to one user exclusively, while each user can occupy multiple sub-bands. The power budget of the BS and the users are indicated by constraints \eqref{DL_P_const} and \eqref{UL_P_const}, respectively, in which $P_{\text{b}}$ and $P_{k}$ are the maximum transmit power of the BS and user $k$. Constraint \eqref{pos_P_const} corresponds to the non-negative transmit power for the BS and users. The justification behind constraints \eqref{Btot_const} and \eqref{Btot_const} are given in \eqref{B_const0} and \eqref{Btot_const0}, respectively. Finally, constraint \eqref{unit_mod_const} ensures unit-modulus phase shift at each IRS element.
	
	It is extremely difficult to obtain the optimal solution to problem \eqref{p1} due to the following reasons. First, the variables $\{\alpha_{k,s}\}$ are integers, making it a mixed-integer problem. Second, the problem contains non-convex constraints, and the optimization variables are heavily coupled. Third, it is difficult to explicitly express the rates due to the existence of integrals and lack of tractable expression of $\kappa\left(f\right)$ for all frequencies within the spectrum of interest.  With these in mind, we first focus on a simplified version of the problem by assuming ESB across all sub-bands, which facilitates tractable optimization. Subsequently, we extend our study to address the more general and practical design strategy with ABS, where the bandwidth of each sub-band can be flexibly adjusted to improve spectral utilization and system performance.
	
	\section{Proposed Solution for \\ Resource Allocation with ESB}
	To tackle the complexity arising from the integral form of the rate expressions in Problem \eqref{p1}, we first consider the resource allocation with ESB in this section. Specifically, the bandwidth of each sub-band is expressed as $B_s=\frac{f_{\text{end}}-f_{\text{start}}-(S-1)B_g}{S}$. 
	However, even with ESB, the optimization variables remain nonlinearly coupled inside the integrand of the logarithmic rate expression, making the problem analytically intractable. To facilitate the  derivation of tractable solutions, we assume perfect SI cancellation at the transceivers. Under this assumption, we proceed to derive an upper bound of the rates, which serves as a surrogate objective to guide heuristic solution development for problem \eqref{p1}. Specifically, by applying Jensen’s inequality to the concave logarithmic function, an upper bound of the DL rate can be obtained \cite{cvx} \begin{equation}\label{jensen}
		R^{\text{d}}_{k,s}\leq B_s\log_2\left(1+\int_{f_s-\frac{B_s}{2}}^{f_s+\frac{B_s}{2}}\frac{p_{k,s}\mathbf{v}^H\mathbf{H}_{k}\left(f\right)\mathbf{v}}{ B_sN_0}\dd f\right)\triangleq \bar{R}^{\text{d}}_{k,s},
	\end{equation}where $\mathbf{H}_{k}\left(f\right)=\hat{\mathbf{h}}_k\left(f\right)\hat{\mathbf{h}}_k^{H}\left(f\right)$ and $\hat{\mathbf{h}}^H_k\left(f\right)=\mathbf{h}^{r\,H}_k\left(f\right)\operatorname{diag}\left(\mathbf{g}^t\left(f\right)\right)$. Denoting $\mathbf{G}_{k}\left(f\right)=\hat{\mathbf{g}}\left(f\right)\hat{\mathbf{g}}^{H}\left(f\right)$ and $\hat{\mathbf{g}}^H\left(f\right)=\mathbf{g}^{r\,H}_k\left(f\right)\operatorname{diag}\left(\mathbf{h}_k^t\left(f\right)\right)$, the UL rate admits a similar upper bound $\bar{R}^{\text{u}}_{k,s}$. These upper bounds serve as tractable surrogates for the original rate expressions. While idealized, this simplification provides valuable insight into system behavior and serves as a useful performance benchmark. The impact of residual SI will be considered in Section IV.


	First of all, to tackle the coupling between continuous variables $\{p_{s},q_{k,s}\}$ and binary variables $\{\alpha_{k,s}\}$, we adopt the big-M formulation.  Define $\tilde{p}=\{\tilde{p}_{k,s}\mid\tilde{p}_{k,s}=\alpha_{k,s}p_{s},k\in\mathcal{K}, s\in\mathcal{S}\}$ and $\tilde{q}=\{\tilde{q}_{k,s}\mid\tilde{q}_{k,s}=\alpha_{k,s}q_{k,s},k\in\mathcal{K}, s\in\mathcal{S}\}$, and impose the following additional constraints:\begin{subequations}\label{BigM}\begin{align}
			&0\leq \tilde{p}_{k,s}\leq \alpha_{k,s}P_{\text{b}},k\in\mathcal{K}, s\in\mathcal{S}\label{BigM_p_d_1}\\
			&p_{s}-\left(1-\alpha_{k,s}\right)P_{\text{b}}\leq\tilde{p}_{k,s}\leq p_{s},k\in\mathcal{K}, s\in\mathcal{S}\label{BigM_p_d_2}\\
			&0\leq\tilde{q}_{k,s}\leq \alpha_{k,s}P_{k},k\in\mathcal{K}, s\in\mathcal{S}\label{BigM_p_u_1}\\
			&q_{k,s}-\left(1-\alpha_{k,s}\right)P_{k}\leq \tilde{q}_{k,s}\leq q_{k,s},k\in\mathcal{K}, s\in\mathcal{S}.\label{BigM_p_u_2}
		\end{align}
	\end{subequations} It is noted that when constraints \eqref{binary} and \eqref{BigM} are satisfied, \eqref{DL_P_const} and \eqref{UL_P_const} can be equivalently transformed into \begin{subequations}\label{P}\begin{align}
			&\sum_{s=1}^{S}\sum_{k=1}^{K}\tilde{p}_{k,s}\leq P_{\text{b}},\label{C1}\\
			&	\sum_{s=1}^{S}\tilde{q}_{k,s}\leq P_{k},k\in\mathcal{K}.\label{C2}
	\end{align}\end{subequations}Also, $\bar{R}^{\text{d}}_{k,s}$ and $\bar{R}^{\text{d}}_{k,s}$ are equivalently written as $
	\bar{R}^{\text{d}}_{k,s}=B_s \log_2\left(1+\int_{f_s-\frac{B_s}{2}}^{f_s+\frac{B_s}{2}}\frac{\tilde{p}_{k,s}\mathbf{v}^H\mathbf{H}_{k}\left(f\right)\mathbf{v}}{B_sN_0}\dd f\right)$ and $
	\bar{R}^{\text{u}}_{k,s}=B_s \log_2\left(1+\int_{f_s-\frac{B_s}{2}}^{f_s+\frac{B_s}{2}}\frac{\tilde{q}_{k,s}\mathbf{v}^H\mathbf{G}_{k}\left(f\right)\mathbf{v}}{B_sN_0}\dd f\right)$. To this end, the problem \eqref{p1} can be simplified into	\begin{subequations}\label{p2}
	\begin{align}
	&\max_{B,\mathbf{v},p,q,\mathbf{\alpha},\tau}\quad   \tau  \\
	&\text{s.t.} \quad \omega_{i,k}\sum_{s=1}^{S}\alpha_{k,s}\bar{R}^{i}_{k,s}\geq \tau, i\in\{\text{d,u}\}, k\in\mathcal{K},\label{rate_const1}\\
	& \quad \quad \eqref{binary}, \eqref{subband_const},\eqref{unit_mod_const},\eqref{BigM},\eqref{P},\eqref{bi2real_1}\nonumber.
	\end{align}
\end{subequations}


	To overcome the difficulty introduced by the binary constraint in \eqref{binary}, we transform it into a combination of two real-valued constraints, formulated as\begin{subequations}\begin{align}
			&0\leq \alpha_{k,s}\leq 1, k\in\mathcal{K}
			,s\in\mathcal{S},\label{bi2real_1}\\
			&\sum_{s=1}^{S}\sum_{k=1}^{K}\left(\alpha_{k,s}-\alpha^{2}_{k,s}\right)\leq 0.\label{bi2real_2}
	\end{align}\end{subequations} Since constraint \eqref{bi2real_2} results in a disconnected feasible region, we incorporate it into the objective function as a penalty term to ensure computational tractability. Thus, problem \eqref{p2} can be approximated as\begin{subequations}\label{p22}
		\begin{align}
			\hspace{-0.3cm}&\max_{\mathbf{v},p,q,\tilde{p},\tilde{q},\mathbf{\alpha},\tau}\quad   \tau- \mu\chi\left(\{\alpha_{k,s}\}\right)\\
			&\text{s.t.} \quad  \eqref{subband_const},\eqref{unit_mod_const},\eqref{BigM},\eqref{P},\eqref{rate_const1},\eqref{bi2real_1}.\nonumber
		\end{align}
	\end{subequations}where $\chi\left(\{\alpha_{k,s}\}\right)\triangleq\sum_{s=1}^{S}\sum_{k=1}^{K}\left(\alpha_{k,s}-\alpha^{2}_{k,s}\right)$ and $\mu\geq 0$ serves as a penalty parameter to penalize the violation of constraint \eqref{bi2real_2}. Notably, to maximize the objective function \eqref{p22} when $\mu\rightarrow\infty$, the optimal $\{\alpha^*_{k,s}\}$ should meet the condition $\sum_{s=1}^{S}\sum_{k=1}^{K}\left(\alpha^*_{k,s}-\alpha^{*\,2}_{k,s}\right)\leq 0$. On the other hand, since $\{\alpha^*_{k,s}\}$ also satisfy constraint \eqref{bi2real_1}, which indicates $\sum_{s=1}^{S}\sum_{k=1}^{K}\left(\alpha^*_{k,s}-\alpha^{*\,2}_{k,s}\right)\geq 0$. Thus, we have $\sum_{s=1}^{S}\sum_{k=1}^{K}\left(\alpha^*_{k,s}-\alpha^{*\,2}_{k,s}\right)=0$ and accordingly $\{\alpha^*_{k,s}\}\in\{0,1\}$ follows, $ k\in\mathcal{K},s\in\mathcal{S}$.
	
	After such manipulation, we observe that the convex term $\alpha^{2}_{k,s}$ makes the objective function non-concave. Together with the unit-modulus modulus constraint, this results in the non-convexity of  problem \eqref{p22}. To address this issue, we adopt the SCA technique. Specifically, the lower bound of $\alpha^{2}_{k,s}$ is obtained as\begin{equation}
		\alpha^{2}_{k,s}\geq -\alpha_{k,s}^{r2}+2\alpha_{k,s}^{r}\alpha_{k,s},
	\end{equation}where $\alpha_{k,s}^{r}$ represent the given feasible points. Then, by relaxing the non-convex unit-modulus constraint, we can approximate problem \eqref{p22} as\begin{subequations}\label{p222}
		\begin{align}
			\hspace{-0.5cm}&\max_{\mathbf{v},p,q,\tilde{p},\tilde{q},\mathbf{\alpha},\tau}\quad   \tau- \mu\chi^{\text{ub}}\left(\{\alpha_{k,s}\}\right)	\\
			&\text{s.t.}  \quad |[\mathbf{v}]_n|\leq1, n\in\mathcal{N},\label{relaxed_unit_mod_const}\\
			&\quad\quad\eqref{subband_const},\eqref{BigM},\eqref{P},\eqref{rate_const1},\eqref{bi2real_1}.\nonumber
		\end{align}
	\end{subequations}where $\chi^{\text{ub}}\left(\{\alpha_{k,s}\}\right)=\sum_{s=1}^{S}\sum_{k=1}^{K}\left(\alpha_{k,s}+\alpha_{k,s}^{r2}-2\alpha_{k,s}^{r}\alpha_{k,s}\right)$.

	
	Then,  we introduce a set of slack variables $\beta^{\text{d}}=\{\beta^{\text{d}}_{k,s},k\in\mathcal{K},s\in\mathcal{S}\}$  such that\begin{equation}\label{beta_d}
		\frac{1}{\beta^{\text{d}}_{k,s}}\leq \tilde{p}_{k,s},k\in\mathcal{K},s\in\mathcal{S}.
	\end{equation} It is noticed that the term $\int_{f_s-\frac{B_s}{2}}^{f_s+\frac{B_s}{2}}\frac{\mathbf{v}^H\mathbf{H}_{k}\left(f\right)\mathbf{v}}{\beta^{\text{d}}_{k,s} B_sN_0}\dd f$ is jointly convex with respect to (w.r.t.) $\beta^{\text{d}}_{k,s}$ and $\mathbf{v}$. Thus, a linear lower bound of it can be obtained as\begin{align}
		\hspace{-0.2cm}&\int_{f_s-\frac{B_s}{2}}^{f_s+\frac{B_s}{2}}\frac{\mathbf{v}^H\mathbf{H}_{k}\left(f\right)\mathbf{v}}{\beta^{\text{d}\,r}_{k,s}B_s N_0}\dd f\geq 2\operatorname{Re}\left\{\int_{f_s-\frac{B_s}{2}}^{f_s+\frac{B_s}{2}}\frac{\mathbf{v}^{r\,H}\mathbf{H}_{k}\left(f\right)\mathbf{v}}{\beta^{\text{d}\,r}_{k,s} B_sN_0}\dd f\right\}\nonumber\\
		&-\int_{f_s-\frac{B_s}{2}}^{f_s+\frac{B_s}{2}}\frac{\mathbf{v}^{r\,H}\mathbf{H}_{k}\left(f\right)\mathbf{v}^r}{\beta^{\text{d}\,r}_{k,s} B_sN_0}\dd f-\int_{f_s-\frac{B_s}{2}}^{f_s+\frac{B_s}{2}}\frac{\mathbf{v}^{r\,H}\mathbf{H}_{k}\left(f\right)\mathbf{v}^r}{\beta^{\text{d}\,r\,2}_{k,s} B_s N_0}\dd f\nonumber\\
		&\times\left(\beta^{\text{d}}_{k,s} -\beta^{\text{d}\,r}_{k,s} \right)\triangleq \mathcal{G}\left(\beta^{\text{d}}_{k,s},\mathbf{v},\mathbf{H}_k\left(f\right)\right),
	\end{align}where $\beta^{\text{d}\,r}_{k,s}$ and $\mathbf{v}^r$ are given feasible points. 	Similarly, the slack variables $\beta^{\text{u}}=\{\beta^{\text{u}}_{k,s},k\in\mathcal{K},s\in\mathcal{S}\}$ for the UL counterpart are introduced to satisfy\begin{equation}
	\label{beta_u}
	\frac{1}{\beta^{\text{u}}_{k,s}}\leq \tilde{q}_{k,s},k\in\mathcal{K},s\in\mathcal{S}.
	\end{equation} Denote $\{\beta=\beta^{i},i\in\{\text{d,u}\}\}$ and problem \eqref{p222} can thus be approximated by\begin{subequations}\label{p2222}
	\begin{align}
		\hspace{-0.6cm}&\max_{\substack{\mathbf{v},p,q,\tilde{p},\tilde{q},\\\mathbf{\alpha},\beta,\tau}}\quad   \tau- \mu\chi^{\text{ub}}\left(\{\alpha_{k,s}\}\right)	\\
		&\text{s.t.} \quad\omega_{i,k}\sum_{s=1}^{S} \log_2\left(1\!+\!\mathcal{G}\left(\beta^{\text{d}}_{k,s},\mathbf{v},\mathbf{H}_k\left(f\right)\right)\right)\!\!\geq \!\tau,k\!\in\!\mathcal{K},\\
		& \quad\quad\omega_{i,k}\sum_{s=1}^{S} \log_2\left(1\!+\!\mathcal{G}\left(\beta^{\text{u}}_{k,s},\mathbf{v},\mathbf{G}_k\left(f\right)\right)\right)\!\!\geq \!\tau,k\!\in\!\mathcal{K},\\
		&\quad\quad\eqref{subband_const},\eqref{BigM},\eqref{P},\eqref{rate_const1},\eqref{bi2real_1},\eqref{relaxed_unit_mod_const},\eqref{beta_d},\eqref{beta_u},\nonumber
	\end{align}
	\end{subequations}which is convex and can be solved by the interior point method \cite{cvx}.

	The overall algorithm for resource allocation with ESB is summarized in Algorithm \ref{alg1}, where $\epsilon_1$ and $\epsilon_2$ are predefined thresholds. We gradually increase the value of the penalty coefficient $\mu$ as $\mu:=c_1 \mu, c_1>1$. The objective value of problem \eqref{p2222} is non-decreasing as the iterations
	proceed, and it is upper bounded by a finite value subject to the limited budget power at the transceivers and the finite size of the IRS. Thus, the proposed algorithm terminates at the
	optimal objective value of problem \eqref{p2222}. Since the obtained $\mathbf{v}$ mat not satisfy the unit-modulus constraint of \eqref{p2}, we reconstruct it as\begin{equation}\label{reconst}
		\left[\hat{\mathbf{v}}\right]_n=\left[\mathbf{v}\right]_n/\left|\left[\mathbf{v}\right]_n\right|,n\in\mathcal{N}.
	\end{equation} Then, by performing the remaining step in   Algorithm \ref{alg1}, we obtain a lower bound to the optimal objective of problem \eqref{p2}. The computational complexity of Algorithm
	\ref{alg1} is given by $\mathcal{O}\left(L_{\text{iter}}\left(N + A\right)^{2.5} A\right)$, where $A=KS + K + S + 1$ and $L_{\text{iter}}$ denotes the  number of iterations required for reaching convergence \cite{complexity}.
	
	
	\begin{algorithm}[t]
		\caption{Resource Allocation with ESB}\label{alg1}
		\begin{algorithmic}[1]
			\STATE \textbf{Initialize}  $\mathbf{v}^0$,  $\{\beta^{u\, 0}_{k,s}\}$, $\{\beta^{d\, 0}_{k,s}\}$, $\mu> 0$, $c_1>1$, and threshold $\epsilon_1$, $\epsilon_2$.
			\REPEAT
			\STATE Set: $r:=0$.
			\REPEAT
			\STATE Solve problem \eqref{p222} to obtain $\mathbf{v}^*$, $\{p_{k,s}^*\}$, $\{q_{k,s}^*\}$, $\{\tilde{p}_{k,s}^*\}$, $\{\tilde{q}_{k,s}^*\}$,$\{\beta^{\text{d}\, *}_{k,s}\}$,  $\{\beta^{\text{u}\, *}_{k,s}\}$,   and $\{\alpha_{k,s}^*\}$.
			\STATE Set: $r:=r+1$.
			\STATE Update:  $\mathbf{v}^r:=\mathbf{v}^*$,  $\beta^{\text{d}\,r}_{k,s}:=\upsilon^{\text{d}\, *}_{k,s}$, $\beta^{\text{u}\,r}_{k,s}:=\upsilon^{\text{u}\, *}_{k,s}$.
			\UNTIL{} the increase of the objective value of problem \eqref{p222} is below $\epsilon_1$.
			\STATE Update: $\mathbf{v}^0:=\mathbf{v}^r$, $\tilde{p}_{k,s}^0:=\tilde{p}_{k,s}^r$,  $\tilde{q}_{k,s}^0:=\tilde{q}_{k,s}^r$, $\upsilon^{\text{u}\, 0}_{k,s}:=\upsilon^{\text{u}\, r}_{k,s}$, $\upsilon^{\text{d}\, 0}_{k,s}:=\upsilon^{\text{d}\, r}_{k,s}$, $\mu:=c_1\mu$.
			\UNTIL{} $\chi\left(\{\alpha_{k,s}\}\right)$ is below $\epsilon_2$.
			\STATE \label{reconstruct}Reconstruct the IRS phase shifts based on \eqref{reconst}.
			\STATE Update $ \{p_{k,s},q_{k,s},\alpha_{k,s},\tau\}$ based on the reconstructed IRS phase-shift vector.
		\end{algorithmic}
	\end{algorithm}
		
	\section{Proposed Solutions for \\ Resource Allocation with ASB}
	For the resource allocation with ASB, the problem is more challenging to solve since the sub-band bandwidths are involved in the optimization. In this section, we extend the proposed solution in Section III to a two-layer penalty-based algorithm to solve problem \eqref{p1}, which includes an inner iteration that solves a penalized optimization problem by applying the BCD method, and an  outer layer that updates the penalty coefficient, until the convergence is achieved.

	\subsection{Convex Approximation of Molecular Absorption Factor}

	The main challenge in solving problem \eqref{p1} falls on the integrals in the channel expressions. Note that although the channel is frequency-dependent, its variation within sub-bands are relatively small when the sub-bands are within a THz TW and have relatively small bandwidths. Thus, we reasonably assume that the channels remain unchanged within each sub-band. Such assumption has also been adopted in \cite{shafie2021spectrum,han2016distance}, which enables the removal of integrals in the channel expressions. 
	However, challenges still remain due to the lack of tractable expression for $\kappa(f)$.  Although the values of molecular absorption coefficients can be obtained from the HITRAN database \cite{HITRAN}, there does not exist a tractable expression that maps $f$ to  $\kappa(f)$. To tackle such intractability, we adopt the convex approximation in \cite{shafie2021spectrum}, which models $\kappa\left(f\right)$ in the spectrum of interest using an exponential function of $f$, given by\begin{equation}\label{convex_approx}
	\hat{k}(f)=e^{\sigma_1+\sigma_2 f}+\sigma_3,
	\end{equation}where $\sigma_1$, $\sigma_2$, and $\sigma_3$ are the fitted curve parameters.

	\subsection{Outer Layer: Update Penalty Coefficient}
	By  dealing with the non-integer variable and relaxing the unit modulus constraint following the same procedures as in the last section, problem \eqref{p1} is approximated by\begin{subequations}\label{p11}
		\begin{align}
			&\max_{\substack{B,\mathbf{v},p,q,\\ \tilde{p},\tilde{q},\mathbf{\alpha},\tau}}\quad   \tau- \mu\chi^{ub}\left(\{\alpha_{k,s}\}\right)	\\
			&\text{s.t.}  \quad \eqref{rate_const}, \eqref{subband_const},\eqref{BigM},\eqref{P},\eqref{bi2real_1},\eqref{B_const},\eqref{Btot_const},\eqref{relaxed_unit_mod_const}.\nonumber
		\end{align}
	\end{subequations}
	
	The difficulty in solving problem \eqref{p11} arises due to the fact that the rate expressions in \eqref{DL_rate} and \eqref{UL_rate} are highly complex w.r.t. to the design variables in $B$. Thus, we approach to solve the problem by considering $f$ as a set of independent design variables, and integrate \eqref{f_B} as an equality constraint. Then, we convert the equality constraint \eqref{f_B} into a quadratic function, which is then also added as a penalty term the objective function of \eqref{p222}, leading to\begin{subequations}\label{p3}
		\begin{align}
			\hspace{-0.5cm}&\max_{\substack{B,f,\mathbf{v},p,q,\\ \tilde{p},\tilde{q},\mathbf{\alpha},\tau}}\quad   \tau- \mu\left(\chi^{\text{ub}}\left(\alpha_{k,s}\right)+\Xi\left(f_s,B_s\right)\right)	\\
			&\text{s.t.}\quad\eqref{rate_const},\eqref{subband_const},\eqref{B_const},\eqref{Btot_const},\eqref{unit_mod_const},\eqref{BigM},\eqref{P},\eqref{bi2real_1},\nonumber
		\end{align}
	\end{subequations}where $\Xi\left(f_s,B_s\right)=\sum_{s=1}^{S}\left|f_s\!-\!\left(f_{\text{start}}+\sum_{i=1}^{s-1}B_{i}+(s\!-\!1)B_g\right.\right.
	\\ \left.\left.+0.5B_{s}\right)\right|^2$.

	\subsection{Inner Layer: BCD Algorithm for Solving Problem \eqref{p3}}
	\subsubsection{Optimizing $f$ for Given $\{B,\mathbf{v},p,q, \tilde{p},\tilde{q},\mathbf{\alpha}\}$}
	Rewrite the cascaded channel gain as $\hat{\mathbf{h}}_{k}^{ H}\left(f_s\right)\mathbf{V}\hat{\mathbf{h}}_{k} \left(f_s\right)$, where $\mathbf{V}=\mathbf{v}\mathbf{v}^H$. Note that $\hat{\mathbf{h}}_{k}^{ H}\left(f_s\right)\mathbf{V}\hat{\mathbf{h}}_{k} \left(f_s\right)$ is convex w.r.t. $\hat{\mathbf{h}}_{k} \left(f\right)$, we can obtain its lower bound  by finding its first-order Taylor expansion as \begin{subequations}\label{channel_f}\begin{align}\hat{\mathbf{h}}_{k} \!\left(f\right)\!\geq 
			& 2\operatorname{Re}\!\left\{\!\hat{\mathbf{h}}_{k} \left(f_s^r\right)^H \!\mathbf{V} \hat{\mathbf{h}}_{k} \left(\!f\!\right)\!\right\}\!-\!\hat{\mathbf{h}}_{k} \left(f^r_s\right)^H \!\mathbf{V}\hat{\mathbf{h}}_{k}\! \left(f_s^r\right)\nonumber\\
			=&\operatorname{Re}\left\{\mathbf{b}_{k,s}^{H}\hat{\mathbf{h}}_{k} \left(f_s\right)\right\}+c^{\text{d}}_{k,s} \\
			=&c^{\text{d}}_{k,s}+\underbrace{\left(\frac{c}{4\pi}\right)^2\frac{1}{\tilde{d}_k}\frac{e^{-\frac{1}{2}\kappa\left(f_s\right)\bar{d}_k}}{f_s^2}}_{\nu_{k,s}\left(f_s\right)}\times\nonumber\\
			\!\!&\!\underbrace{\operatorname{Re}\!\left\{\!\sum_{n_x=1}^{N_x}\sum_{n_y=1}^{N_y}\!\left| b^{\text{d},k,s}_{n_x,n_y}\right|\!e^{-j\left(\pi f_s\frac{\lambda_c}{c}\!\left(\!\delta^k_{n_x}+\delta^k_{n_y}\!\right)\!-\angle b^{\text{d},k,s}_{n_x,n_y}\!\right)\!}\!\right\}}_{\eta^{\text{d}}_{k,s}\left(f_s\right)}\\
			\triangleq&\gamma_{k,s}^{\text{d}}\left(f_s\right),
	\end{align}	\end{subequations}where $c^{\text{d}}_{k,s}=-\hat{\mathbf{h}}_{k}\left(f_s^r\right)^H \mathbf{V}\hat{\mathbf{h}}_{k}\left(f_s^r\right)$, $\bar{d}_k=d_k+d_{\text{\text{b}}}$, $\tilde{d}_k=d_{\text{\text{b}}}d_k$, $\delta^k_{n_x}=n_x\left(\sin(\psi_r^r)\sin(\varphi_r^r)- \sin(\psi_k^t)\sin(\varphi_k^t)\right)$, $\delta^k_{n_y}=n_y\left(\cos(\varphi_r^r)- \cos(\varphi_k^t)\right)$,  $\mathbf{b}^{\text{d}}_{k,s}\triangleq 2\mathbf{V}^H\hat{\mathbf{h}}_{k}\left(f_s^r\right)$, and the $\left(n_x,n_y\right)$-th entry of $\mathbf{b}^{\text{d}}_{k,s}$ is denoted as $\left|b^{\text{d},k,s}_{n_x,n_y}\right|e^{j\angle{b^{\text{d},k,s}_{n_x,n_y}}}$. Thus, $\eta_{k,s}^{\text{d}}\left(f_s\right)$ can be written as
	\begin{equation}\hspace{-0.1cm}
		\eta_{k,s}^{\text{d}}\!\left(f_s\right)\!=\!\sum_{n_x=1}^{N_x}\!\sum_{n_y=1}^{N_y}\!\left| b^{\text{d},k,s}_{n_x,n_y}\right|\!\cos\!\left(\!\pi f_s\frac{\lambda_c}{c}\!\left(\!\delta^k_{n_x}\!+\!\delta^k_{n_y}\!\right)\!-\!\angle b^{\text{d},k,s}_{n_x,n_y}\!\right)\!.
	\end{equation}
	The lower bound $\gamma_{k,s}^{\text{u}}\left(f_s\right)$ for UL cascaded channel, $\eta^{\text{u}}_{k,s}\left(f_s\right)$, and $c^{\text{u}}_{k,s}$ are similarly defined.
	
	It is noted that $\eta^i_{k,s}\left(f_s\right),i\in\{\text{d,u}\},$ is still neither convex nor concave w.r.t. $f_s$.  We attempt to construct a surrogate function that locally approximates the objective function by using the second-order Taylor expansion.  Denote $\nabla \eta^i_{k,s}\left(f_s^r\right)$ and $\nabla^2 \eta^i_{k,s}\left(f_s^r\right)$  as the gradient and the Hessian of $\eta^i_{k,s}\left(f_s\right)$ over $f_s$, given respectively by \eqref{gradient} and \eqref{hessian}, shown at the top of next page. Based on \eqref{hessian}, we select a positive real number that satisfies $\xi^i_{k,s}\geq \eta^i_{k,s}\left(f_s\right)$, with the closed-form expression given by $\xi^i_{k,s}=\left(\pi \frac{\lambda_c}{c}\right)^2\sum_{n_x=1}^{N_x}\sum_{n_y=1}^{N_y}\left| b^{i,k,s}_{n_x,n_y}\right|\left(2n_x+2n_y\right)^2$. Then, with given feasible point $f_s^r$, a global lower bound of $\eta^{i}_{k,s}\left(f_s\right)$ can be obtained as\begin{subequations}\begin{align}
			\eta^{i}_{k,s}\left(f_s\right)\geq &\eta_{k,s}^i\left(f_s^r\right)\!+\!\nabla \eta_{k,s}^i\left(f_s^r\right)\left(f_s\!-\!f_s^r\right)  \!-\!\frac{\xi^i_{k,s}}{2}\left(f_s\!-\!f_s^r\right)^2\!,\\
			=&-\frac{\xi^i_{k,s}}{2}f_s^2+(\nabla \eta^{i}_{k,s}\left(f_s^r\right)+\xi^i_{k,s} f_s^r)f_s\nonumber\\
			&+\eta^{i}_{k,s}\left(f_s^r\right)-\nabla \eta^{i}_{k,s}\left(f_s^r\right)f_s^r-\frac{\xi^i_{k,s}}{2}f_s^{r 2},\\
			\triangleq& \tilde{\eta}^{i}_{k,s}\left(f_s\right), i\in\{\text{d,u}\}, k\in\mathcal{K}, s\in\mathcal{S},
		\end{align}
	\end{subequations}   which is concave w.r.t. $f_s$. 
	
	\begin{figure*}[!t]
		\begin{align}\label{gradient}
			\nabla \eta^i_{k,s}\left(f_s\right)=&-\pi \frac{\lambda_c}{c}\sum_{n_x=1}^{N_x}\sum_{n_y=1}^{N_y}\left| b^{i,k,s}_{n_x,n_y}\right|\left(\delta_{n_x}^k+\delta_{n_y}^k\right)\sin\left(\pi f_s\frac{\lambda_c}{c}\left(\delta_{n_x}^k+\delta_{n_y}^k\right)-\angle b^{i,k,s}_{n_x,n_y}\right).
		\end{align}

		\begin{align}	\label{hessian}
		\nabla^2\eta^i_{k,s} \left(f_s\right)=&-\left(\pi \frac{\lambda_c}{c}\right)^2\sum_{n_x=1}^{N_x}\sum_{n_y=1}^{N_y}\left| b^{i,k,s}_{n_x,n_y}\right|\left(\delta_{n_x}^k+\delta_{n_y}^k\right)^2\cos\left(\pi f_s\frac{\lambda_c}{c}\left(\delta_{n_x}^k+\delta_{n_y}^k\right)-\angle b^{i,k,s}_{n_x,n_y}\right).
		\end{align}
	\end{figure*}

	The path gain of LSI channel is given by $\beta_{\text{LSI}}\left(d\right)=\frac{1}{4}\left(\frac{1}{\left(\kappa_{\lambda }d\right)^2}-\frac{1}{\left(\kappa_{\lambda }d\right)^4}+\frac{1}{\left(\kappa_{\lambda} d\right)^6}\right)$, where $\kappa_{\lambda}=\frac{2\pi}{\lambda}$ is the wave number and $\lambda$ is the corresponding wavelength \cite{SI}. Since the wavelength is small in THz band, we assume $\beta_{\text{LSI}}\left(d\right)\approx \frac{1}{4}\left(\frac{1}{\left(\kappa_{\lambda }d\right)^2}\right)$. Then, we can denote the channel gain of the $k$-th user's LSI channel as $\nu_{0,k,s}\left(f_s\right)= \left(\frac{c}{4\pi f_s }\right)^2\left(\frac{1}{\tilde{d}_{0,k}}\right)e^{-\frac{1}{2}\kappa\left(f_s\right)\bar{d}_{0,k}}$, where $\bar{d}_{0,k}=2d_{0,k}$ and $\tilde{d}_{0,k}=d_{0,k}^2$. Similarly, that of the BS's LSI channel is defined as $\nu_{0,K+1,s}\left(f_s\right)\triangleq \left(\frac{c}{4\pi f_s }\right)^2\left(\frac{1}{\tilde{d}_{0}}\right)e^{-\frac{1}{2}\kappa\left(f_s\right)\bar{d}_{0}}$, where $\bar{d}_{0,K+1}=2d_0$ and $\tilde{d}_{0,K+1}=d_0^2$. To handle the non-convexity of the rate expressions, we first introduce two sets of slack variables $x=\{x^{i}_{k,s},i\in\{\text{d,u}\},k\in\mathcal{K},s\in\mathcal{S}\}$ and $y=\{y_{k,s},k\in\mathcal{K}'=\{1,\dots,K+1\},s\in\mathcal{S}\}$, which respectively satisfy
	\begin{align}\label{x_const1}
		2^{x_{k,s}^{\text{d}}}\leq&\,\Gamma \tilde{q}_{k,s}\nu_{0,k,s}\left(f_s\right)+\tilde{p}_{k,s}\nu_{k,s}\left(f_s\right)\tilde{\eta}^d_{k,s}\left(f_s\right)\nonumber\\
		&+\tilde{p}_{k,s}c^{\text{d}}_{k,s}+B_sN_0, k\in\mathcal{K},s\in\mathcal{S},
	\end{align}
	\begin{align}\label{x_const2}
		2^{x_{k,s}^{\text{u}}}\leq&\,\Gamma \tilde{p}_{k,s}\nu_{0,K+1,s}\left(f_s\right)+\tilde{q}_{k,s}\nu_{k,s}\left(f_s\right)\tilde{\eta}^u_{k,s}\left(f_s\right)\nonumber\\
		&+\tilde{q}_{k,s}c^{\text{u}}_{k,s}+B_sN_0, k\in\mathcal{K},s\in\mathcal{S},
	\end{align}and
	\begin{equation}\label{y_const}
		y_{k,s}\geq \nu_{0,k,s}\left(f_s\right),k\in\mathcal{K}'.
	\end{equation}Then, we can re-express the DL rate as
	$\tilde{R}_{k,s}^{\text{d}}= B_sx_{k,s}^{\text{d}}-B_s\log_2\left(\Gamma \tilde{q}_{k,s}y_{s}+B_sN_0\right)$. Since such expression is convex, we obtain its lower bound by applying the SCA technique as
	\begin{align}
		&\hspace{-0.2cm}\tilde{R}_{k,s}^{\text{d}}
		\geq B_sx_{k,s}^{\text{d}}
		-B_s\Big(\!\log_2\left(\Gamma \tilde{q}_{k,s}y_{k,s}^r\!+\!B_sN_0\right)\nonumber\\
		&+\frac{\frac{1}{\ln2}\Gamma\tilde{q}_{k,s}}{\Gamma\tilde{q}_{k,s}y_{k,s}^r\!+\!B_sN_0}\left(y_{k,s}\!-\!y_{k,s}^r\right)\!\Big)\!
		\triangleq\tilde{\tilde{R}}^{\text{d}}_{k,s}\left(x^{\text{d}}_{k,s},y_{k,s}\right),
	\end{align}
	where $y_{k,s}^r$ is given feasible point. The lower bound of $R_{k,s}^{\text{u}}$ is similarly defined as $\tilde{\tilde{R}}^{\text{u}}_{k,s}\left(x_{k,s}^{\text{u}},y_{k,s}\right)\triangleq B_s x_{k,s}^{\text{u}}-B_s\times\\
	\!\!\!\left(\!\log_2\!\left(\Gamma \tilde{p}_{k,s}y_{K+1,s}^r\!\!+\!B_sN_0\right)\! +\!\frac{\frac{1}{\ln2}\Gamma\tilde{p}_{k,s}\left(y_{K\!+\!1,s}- y_{K\!+\!1,s}^r\right)}{\Gamma\tilde{p}_{k,s}y_{K+1,s}^r+B_s\!N_0\!}\!\right)$.
	
	To deal with the non-convex constraint \eqref{y_const}, we further introduce two sets of slack variables $u=\{u_{s},s\in\mathcal{S}\}$ and $w=\{w_s,s\in\mathcal{S}\}$, satisfying $
	\sigma_1+\sigma_2f_s\leq \log\left(u_s\right)$ and 
	$\sigma_1+\sigma_2f_s\geq\log\left(w_s\right)$,
	where the RHS of the latter can be found as $\log\left(w^r_s\right)+\frac{1}{w^r_s}\left(w_s-w_s^r\right)$ and $w^r_s$ is given feasible point. Then, constraint \eqref{y_const} can be approximated by \begin{equation}\label{v_ub}
		y_{k,s}\geq \left(\frac{c}{4\pi f_s}\right)^2\frac{1}{\tilde{d}_{0,k}}e^{-\frac{1}{2}\left(w_s+\sigma_3\right)\bar{d}_{0,k}},k\in\mathcal{K}'.
	\end{equation}
	
	To deal with non-convex constraint \eqref{x_const1}, we re-express it as\begin{align}\label{34}
		&\hspace{-0.2cm}\frac{2^{x_{k,s}^{\text{d}}}\!-\!\left(\!\tilde{p}_{k,s}c^{\text{d}}_{k,s}\!+\!B_sN_0\!\right)}{e^{-\frac{1}{2}\kappa(\!f_s\!)\bar{d}_k}}\!-\! \Gamma\tilde{q}_{k,s}\!\left(\!\frac{c}{4\pi f_s}\!\right)^2\!\!\frac{1}{\tilde{d}_{0,k}}e^{-\frac{1}{2}\kappa(f_s)\left(\!\bar{d}_{0,k}-\bar{d}_k\!\right)}\nonumber\\
		&\leq\tilde{p}_{k,s}\left(\frac{c}{4\pi f_s}\right)^2\frac{1}{\tilde{d}_k}\tilde{\eta}^{\text{d}}_{k,s}\left(f_s\right),k\in\mathcal{K},s\in\mathcal{S}.
	\end{align} Then, we introduce another set of slack variable $z=\{z^i_{k,s},i\in\{\text{d,u}\},k\in\mathcal{K},s\in\mathcal{S}\}$, satisfying  \begin{align}\label{39}
		&\frac{1}{z_{k,s}^{i}}\geq \frac{2^{x_{k,s}^{i}}}{e^{-\frac{1}{2}\left(u_s+\sigma_3\right)\bar{d}_{k}}},i\in\{\text{d,u}\},k\in\mathcal{K},s\in\mathcal{S}.
	\end{align}By finding the lower bound of $\frac{e^{-\frac{1}{2}\left(u_s+\sigma_3\right)\bar{d}_{k}}}{2^{x_{k,s}^{\text{d}}}}$, we can approximate \eqref{39} by\begin{align}\label{z_const}
		&\frac{e^{-\frac{1}{2}\left(u_s^r+\sigma_3\right)\bar{d}_{k}}}{2^{x_{k,s}^{\text{d}\,r}}} - \frac{\frac{1}{2}\bar{d}_{k} e^{-\frac{1}{2}\left(u_s^r+\sigma_3\right)\bar{d}_{k}}}{2^{x_{k,s}^{i\,r}}}(u_s - u_s^r)\nonumber \\
		&- \frac{e^{-\frac{1}{2}\left(u_s^r+\sigma_3\right)\bar{d}_{k}} \ln(2)}{2^{x_{k,s}^{\text{d}\,r}}}(x_{k,s}^{\text{d}} - x_{k,s}^{\text{d}\,r})\geq z_{k,s}^{\text{d}},
	\end{align}where $u_s^r$ and $x_{k,s}^{\text{d}\,r}$ are given feasible points. By further introducing two sets of slack variables $t=\{t_{k,s},k\in\mathcal{K},s\in\mathcal{S}\}$ and $r=\{r_{k,s},k\in\mathcal{K},s\in\mathcal{S}\}$, satisfying $t_{k,s} \leq e^{-\frac{1}{2}\kappa\left(f_s\right)\bar{d}_k}$ and $r_{k,s}\geq  \frac{e^{-\frac{1}{2}\left(w_s+\sigma_3\right)\bar{d}_k}}{f_s^2}$, we can rewrite constraint \eqref{34} as
	\begin{align}\label{42}
		&\frac{f_s^2}{z^{\text{d}}_{k,s}}-\frac{\tilde{p}_{k,s}c^{\text{d}}_{k,s}f_s^2}{t_{k,s}}-\frac{B_sN_0}{r_{k,s}}-\Gamma\tilde{q}_{k,s}\left(\frac{c}{4\pi}\right)^2\frac{1}{\tilde{d}_{0,k}}\times\nonumber\\
		&\!\!\left(\!e^{-\frac{1}{2}\left(u_s^r+\sigma_3\right)\left(\bar{d}_{0,k}-\bar{d}_k\right)}\right)
		\!\!\leq\!\tilde{p}_{k,s}\left(\frac{c}{4\pi }\right)^2\!\!\frac{1}{\tilde{d}_k}\tilde{\eta}^{\text{d}}_{k,s}\left(f_s\right),k\in\mathcal{K},s\in\mathcal{S}.
	\end{align}Since the left-hand-side (LHS) of \eqref{42} is concave w.r.t. $u_s$, we construct its upper bound by finding its first-order Taylor expansion and transform \eqref{42} into \eqref{30_convert}, also shown at the top of next page, where $u_s^r$ is given feasible point.  Similarly, constraint \eqref{x_const2} can be transformed into \eqref{31_convert}, shown at the top of next page.


	To this end, the subproblem can be approximated by\begin{subequations}\label{sub1}
		\begin{align}\hspace{-0.4cm}	
			&\max_{\substack{x,y,u,w,z,\\t,r, f,\tau}} \quad   \tau- \mu\left(\chi^{\text{ub}}\left(\alpha_{k,s}\right)+\Xi\left(f_s,B_s\right)\right)\\
			&\text{s.t.} \quad  \omega_{i,k}\sum_{s=1}^{S}\tilde{\tilde{R}}_{k,s}^{i}\left(x_{k,s}^{i},y_{k,s}\right)\geq t,i\in\{\text{d,u}\},k\in\mathcal{K},\\
			& \quad\quad\sigma_1+\sigma_2f_s\leq \log\left(u_s\right),s\in\mathcal{S},\\
			& \quad\quad\sigma_1+\sigma_2f_s\geq\log\left(w^r_s\right)+\frac{1}{w^r_s}\left(w_s-w_s^r\right),s\in\mathcal{S},\\
			&\quad\quad t_{k,s} \leq e^{-\frac{1}{2}\kappa\left(f_s\right)\bar{d}_k},k\in\mathcal{K},s\in\mathcal{S},\\
			& \quad\quad r_{k,s}\geq  \frac{e^{-\frac{1}{2}\left(w_s+\sigma_3\right)\bar{d}_k}}{f_s^2},k\in\mathcal{K},s\in\mathcal{S},\\
			& \quad\quad \eqref{v_ub},\eqref{z_const},\eqref{30_convert},\eqref{31_convert},\nonumber
		\end{align}
	\end{subequations}which is convex and can be solved by interior-point method \cite{cvx}. \looseness=-1
	
	\subsubsection{Optimizing $B$ for $\{f,\mathbf{v},p,q, \tilde{p},\tilde{q},\mathbf{\alpha}\}$}
	The subproblem for optimizing $B$ is given by\begin{subequations}\label{sub2}
		\begin{align}
			&\max_{B,\tau} \quad  \tau- \mu\left(\chi^{\text{ub}}\left(\alpha_{k,s}\right)+\Xi\left(f_s,B_s\right)\right)\\
			&\text{s.t.}\quad  \eqref{rate_const},\eqref{B_const},\nonumber
		\end{align}
	\end{subequations} which is convex and can be solved by interior-point method \cite{cvx}. \looseness=-1
	
	\subsubsection{Optimizing $\{\mathbf{v},p,q,\tilde{p},\tilde{q}\}$ for given $\{B,f\}$}
	The subproblem for optimizing $\{\mathbf{v},p,q,\tilde{p},\tilde{q},\alpha\}$ is given by\begin{subequations}\label{sub3}
		\begin{align}
			&\max_{\substack{\mathbf{v},p,q,\\
					\tilde{p},\tilde{q},\alpha,\tau}} \quad  \tau- \mu\left(\chi^{\text{ub}}\left(\alpha_{k,s}\right)+\Xi\left(f_s,B_s\right)\right) \\
			&\text{s.t.} \quad\eqref{rate_const},\eqref{subband_const},\eqref{BigM},\eqref{P},\eqref{bi2real_1},\eqref{relaxed_unit_mod_const}\nonumber.
		\end{align}
	\end{subequations}	We start by introducing a set of slack variables $\varrho^{\text{d}}=\{\varrho^{\text{d}}_{k,s},k\in\mathcal{K},s\in\mathcal{S}\}$  such that\begin{equation}\label{23a}
	\varrho^{\text{d}}_{k,s}\leq \frac{\tilde{p}_{k,s}\mathbf{v}^H\mathbf{H}_{k}\left(f_s\right)\mathbf{v}}{ \Gamma\tilde{q}_{k,s}\left|\mathbf{h}_k\left(f_s\right)\right|^2+B_{s}N_0},k\in\mathcal{K},s\in\mathcal{S}.
	\end{equation}Then, the DL rate constraint can be rewritten as\begin{equation}\label{DLrate}
	\omega_{\text{d},k}\sum_{s=1}^{S}B_s\log_2\left(1+\varrho_{k,s}^{\text{d}}\right)\geq \tau,k\in\mathcal{K}.
	\end{equation}To deal with non-convex constraint \eqref{23a}, we further introduce a set of slack variables  $\upsilon^{\text{d}}=\{\upsilon^{\text{d}}_{k,s},k\in\mathcal{K},s\in\mathcal{S}\}$ that satisfy\begin{equation}\label{23b}
	\upsilon_{k,s}^{\text{d}} \leq \mathbf{v}^H\mathbf{H}_{k}\left(f_s\right)\mathbf{v}, k\in\mathcal{K},s\in\mathcal{S}.
	\end{equation}Thus, constraint \eqref{23a} can be rewritten as\begin{equation}\label{26}
	\varrho^{\text{d}}_{k,s}\left(\Gamma\tilde{q}_{k,s}\left|\mathbf{h}_k\left(f_s\right)\right|^2+B_{s}N_0\right)\leq \tilde{p}_{k,s}\upsilon_{k,s}^{\text{d}},k\in\mathcal{K},s\in\mathcal{S}.
	\end{equation} Note that constraint \eqref{26}  is still non-convex due to the coupled variables. To deal with it, we first rewrite the expression $ab$ as $\frac{1}{4}\left(\left(a+b\right)^2-\left(a-b\right)^2\right)$. Then, we define the lower bound of $\left(a+b\right)^2$ as $\zeta_{\text{lb}}\left(a,b\right)\triangleq\left(a^r+b^r\right)^2+2\left(a^r+b^r\right)\left(a-a^r\right)+2\left(a^r+b^r\right)\left(b-b^r\right)$, where $a^r$ and $b^r$ are given feasible points. On the other hand, we define the upper bound of the expression $ab$ as $\zeta_{\text{ub}}\left(a,b\right)\triangleq\frac{1}{2\varepsilon^r}a^2+\frac{\varepsilon^r}{2}b^2$, where $\varepsilon^r=\frac{a}{b}$. As such, we can re-express constraint \eqref{26} as \begin{align}\label{26_eq}
	&\Gamma\left|\mathbf{h}_k\left(f_s\right)\right|^2\zeta_{\text{ub}}\left(\varrho^{\text{d}}_{k,s},\tilde{q}_{k,s}\right)+B_{s}N_0\varrho^{\text{d}}_{k,s}\nonumber\\
	\leq& \frac{1}{4}\left(\zeta_{\text{lb}}\left(\tilde{p}_{k,s},\upsilon^{\text{d}}_{k,s}\right)\!-\!\left(\tilde{p}_{k,s}-\upsilon^{\text{d}}_{k,s}\right)^2\right),k\in\mathcal{K},s\in\mathcal{S}.
	\end{align}In addition, for the non-convex term $\mathbf{v}^H\mathbf{H}_{k}\left(f_s\right)\mathbf{v}$ in the right-hand-side (RHS) of \eqref{23b}, a lower bound can also be obtained as\begin{align}
	\mathbf{v}^H\mathbf{H}_{k}\left(f_s\right)\mathbf{v}&\geq 2 \operatorname{Re}\left\{\mathbf{v}^{H} \mathbf{H}_{k}\left(f_s\right) \mathbf{v}^r\right\}-\mathbf{v}^{r\,H} \mathbf{H}_{k}\left(f_s\right) \mathbf{v}^r \nonumber\\
	&\triangleq \lambda_{\text{lb}}\left(\mathbf{v},\mathbf{H}_{k}\left(f_s\right)\right).
	\end{align} 
	
	Similarly, by introducing slack variables $\varrho^{\text{u}}=\{\varrho^{\text{u}}_{k,s},k\in\mathcal{K},s\in\mathcal{S}\}$ and $\upsilon^{\text{u}}=\{\upsilon^{\text{u}}_{k,s},k\in\mathcal{K},s\in\mathcal{S}\}$, we can rewrite the UL rate constraint  as\begin{equation}\label{ULrate}
	\omega_{\text{u},k}\sum_{s=1}^{S}B_s\log_2\left(1+\varrho^{\text{u}}_{k,s}\right)\geq \tau,k\in\mathcal{K},
	\end{equation}with the following additional constraints:\begin{equation}
	\varrho^{{\text{u}}}_{k,s}\leq\frac{\tilde{q}_{k,s}\mathbf{v}^H\mathbf{G}_{k}\left(f_s\right)\mathbf{v}}{\Gamma \tilde{p}_{k,s}\left|\mathbf{g}\left(f_s\right)\right|^2+B_{s}N_0}, k\in\mathcal{K},s\in\mathcal{S},
	\end{equation}\begin{equation}
	\upsilon_{k,s}^{\text{u}} \leq \mathbf{v}^H\mathbf{G}_{k}\left(f_s\right)\mathbf{v},  k\in\mathcal{K},s\in\mathcal{S},
	\end{equation} 
	
	Denote $\varrho=\{\varrho^i,i\in\{\text{d,u}\}\}$ and $\upsilon=\{\upsilon^i,i\in\{\text{d,u}\}\}$. To this end, problem \eqref{sub3} can be approximated by	
	\begin{subequations}\label{sub32}
	\begin{align}
		\hspace{-0.5cm}&\max_{\substack{\mathbf{v},p,q,\tilde{p},\tilde{q},\\
				\beta,\upsilon,\alpha,\tau}}\quad   \tau- \mu\chi^{\text{ub}}\left(\{\alpha_{k,s}\}\right)	\\
		&\text{s.t.}  \quad \!\!	\upsilon_{k,s}^{\text{d}} \!\leq\! \lambda_{\text{lb}}\left(\mathbf{v},\mathbf{H}_{k}\left(f_s\right)\right),\upsilon_{k,s}^{\text{u}} \!\leq\! \lambda_{\text{lb}}\left(\mathbf{v},\mathbf{G}_{k}\left(f_s\right)\right),\nonumber\\
		&\quad\quad \,k\in\mathcal{K},s\in\mathcal{S},\label{33b}\\
		&\quad\quad\Gamma\left|\mathbf{g}\left(f_s\right)\right|^2\zeta_{\text{ub}}\left(\beta^{\text{u}}_{k,s},\tilde{p}_{k,s}\right)+B_{s}N_0\beta^{\text{u}}_{k,s}\nonumber\\
		&\quad\quad\,\leq \frac{1}{4}\left(\zeta_{\text{lb}}\left(\tilde{q}_{k,s},\upsilon^{\text{u}}_{k,s}\right)-\left(\tilde{q}_{k,s}-\upsilon^{\text{u}}_{k,s}\right)^2\right),k\in\mathcal{K},s\in\mathcal{S},\label{33d}\\
		&\quad\quad\eqref{subband_const},\eqref{BigM},\eqref{P},\eqref{bi2real_1},\eqref{relaxed_unit_mod_const},\eqref{DLrate},\eqref{26_eq},\eqref{ULrate},\nonumber
	\end{align}
	\end{subequations}which is convex and can be solved by the interior point method \cite{cvx}.

	\subsection{Convergence and Computational Complexity}
	We define an indicator $\rho$  as \begin{equation}
		\rho=\max\{\chi^{\text{ub}}\left(\alpha_{k,s}\right),\Xi\left(f_s,B_s\right)\}.
	\end{equation} In the outer layer, the proposed algorithm the gradually increases $\mu$  until $\rho\leq \epsilon_1$, while in the inner layer, the proposed algorithm iteratively solves problems \eqref{sub1}, \eqref{sub2}, and \eqref{sub3} until the increase in the objective function of \eqref{p3} is less than $\epsilon_2$. The overall algorithm for resource allocation with ASB is summarized in Algorithm \ref{algorithm1}.  The computational complexity of Algorithm \ref{algorithm1} is $\mathcal{O}\left(L_{\text{inn}}L_{\text{out}}\left(\sqrt{C+K}C\left(C+K^2S^2 + K^2S + KS^2\right)\right.\right.\\
	\left.\left.+\left(S+K\right)^{3.5}+\left(N + A\right)^{2.5} A\right)\right)$, where $C=KS+S+1$. $L_{\text{inn}}$ and $L_{\text{out}}$ denote the numbers of iterations required for reaching convergence in the inner layer and outer layer, respectively.


		
		\begin{figure*}
			
			\begin{align}\label{30_convert}
				&\frac{f_s^2}{z^{\text{d}}_{k,s}}-\frac{\tilde{p}_{k,s}c^{\text{d}}_{k,s}f_s^2}{t_{k,s}}-\frac{B_sN_0}{r_{k,s}}-\Gamma\tilde{q}_{k,s}\left(\frac{c}{4\pi}\right)^2\frac{1}{\tilde{d}_{0,k}}\times\left(\!e^{-\frac{1}{2}\left(u_s^r+\sigma_3\right)\left(\bar{d}_{0,k}-\bar{d}_k\right)}-\frac{1}{2} e^{-\frac{1}{2}\left(u_s^r+\sigma_3\right)\left(\bar{d}_{0,k}-\bar{d}_k\right)}\left(u_s-u_s^r\right)\right)
				\nonumber\\
				&\quad\quad
				\leq\tilde{p}_{k,s}\left(\frac{c}{4\pi }\right)^2\frac{1}{\tilde{d}_k}\tilde{\eta}^{\text{d}}_{k,s}\left(f_s\right),k\in\mathcal{K},s\in\mathcal{S}.
			\end{align}
			\begin{align}\label{31_convert}
				&\frac{f_s^2}{z^{\text{u}}_{k,s}}-\frac{\tilde{q}_{k,s}c^{\text{u}}_{k,s}f_s^2}{t_s}-\frac{B_sN_0}{r_s}-\Gamma\tilde{p}_{k,s}\left(\frac{c}{4\pi}\right)^2\frac{1}{\tilde{d}_{0,K+1}}\times\left(\!e^{-\frac{1}{2}\left(u_s^r+\sigma_3\right)\left(\bar{d}_{0,K+1}-\bar{d}_k\right)}-\frac{1}{2} e^{-\frac{1}{2}\left(u_s^r+\sigma_3\right)\left(\bar{d}_{0,K+1}-\bar{d}_k\right)}\left(u_s-u_s^r\right)\right)\nonumber\\
				&\quad\quad
				\leq\tilde{q}_{k,s}\left(\frac{c}{4\pi }\right)^2\frac{1}{\tilde{d}_k}\tilde{\eta}^{\text{u}}_{k,s}\left(f_s\right),k\in\mathcal{K},s\in\mathcal{S}.
			\end{align}
		\end{figure*}

		\begin{algorithm}[t]
			\caption{Resource Allocation with ASB}\label{algorithm1}
			\begin{algorithmic}[1]
				\STATE \textbf{Initialize}  $\mathbf{v}^0$, $\{\tilde{p}_{k,s}^0\}$,  $\{\tilde{q}_{k,s}^0\}$, $\{\upsilon^{u\, 0}_{k,s}\}$, $\{\upsilon^{d\, 0}_{k,s}\}$, $\mu> 0$, $c_1>1$, and threshold $\epsilon_1$, $\epsilon_2$.
				\STATE \textbf{repeat: outer layer} 
				\STATE \quad Set: $r:=0$.
				\STATE \quad \textbf{repeat: inner layer}  
				\STATE \quad Update $f$ by solving problem \eqref{sub1}.
				\STATE \quad Update $B$ by solving problem \eqref{sub2}.
				\STATE \quad Update $\{\mathbf{v},p,q,\tilde{p},\tilde{q},\alpha\}$ by solving problem \eqref{sub32}.
				\STATE \quad \textbf{until}  the increase of the objective value of problem \eqref{p11} \\  \quad is below $\epsilon_1$.
				\STATE \quad Update: $\mu:=c_1\mu$.
				\STATE \textbf{until}  $\rho$ is below $\epsilon_2$.
				\STATE \label{reconstruct}Reconstruct the IRS phase shifts based on \eqref{reconst}.
				\STATE Update $ \{B,f,p,q,\alpha,\tau\}$ based on the reconstructed IRS phase-shift vector.
			\end{algorithmic}
		\end{algorithm}

	\section{Numerical Results}
\begin{figure}[t]
	\centering
	\includegraphics[width=2.7in]{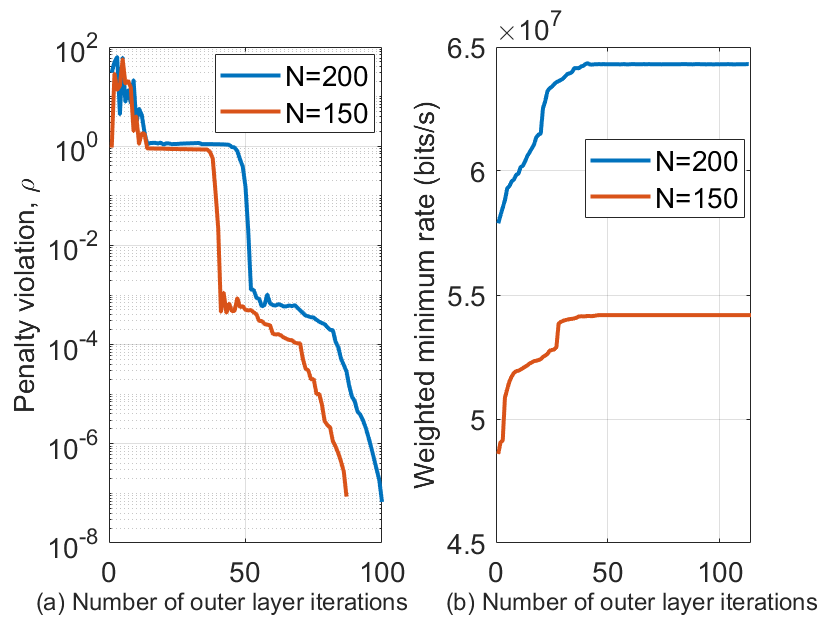}
	\vspace{-0.1cm}
	\caption{One snapshot of constraint violation and convergence behavior of Algorithm \ref{algorithm1}.}
	\label{convergence}
	\vspace{-0.3cm}
\end{figure}

	In this section, numerical results are presented to  evaluate the performance of the proposed schemes. We consider a three-dimensional coordinate system, where the IRS is positioned at $(0,0,0)$ meters (m), the BS is located at $(0,0,-2)$ m, and users are uniformly distributed within a $2$ m radius circle centered at $(0,4,-2.5)$ m. It is assumed that  IRS elements are separated by half of the wavelength, i.e., $\Delta_d=\frac{\lambda_c}{2}$.  The maximum transmit power at the BS and users is set to \(P_{\text{b}}=40~\mathrm{dBm}\) and \(P_{k}=30~\mathrm{dBm}\), respectively. We assume that the antenna separation is \(d_{0}=40~\mathrm{cm}\) and \(d_{0,k}=20~\mathrm{cm}\)  at the BS and users, respectively. Without otherwise specified, the other parameters are set as follows: \(f_{\text{min}}=1.035~\mathrm{THz}\), \(f_{\text{max}}=1.055~\mathrm{THz}\), \(N=10\times 10\), \(K=3\), \(S=6\),  \(B_{\text{max}}=5~\mathrm{GHz}\), \(B_g=0.75~\mathrm{GHz}\), \(\sigma_1=-353.5359\), \(\sigma_2=3.308222\times10^{-10}\), and \(\sigma_3=0.1818\), $\Gamma=-65~\mathrm{dB}$, $N_0=-174~\mathrm{dBm/Hz}$, $c_1=1.25$, $\epsilon_1=10^{-7}$, and $\epsilon_2=10^{-4}$. The constraint violation and convergence behavior of Algorithm \ref{algorithm1} for one snapshot are shown in Fig. \ref{convergence}. It is observed in Fig. \ref{convergence}(a) that the constraint violation indicator, $\rho$, can eventually decrease to a predefined accuracy, $10^{-7}$, after about $48$ iterations for $N = 200$, which indicates that the  constraints \eqref{f_B} and \eqref{bi2real_2} can eventually be  satisfied for problem \eqref{p1}. As such, the two-layer penalty-based algorithm is guaranteed to converge finally. This can also be observed more clearly in Fig. \ref{convergence}(b), where the  objective values of \eqref{p1} obtained by different $N$  increase rapidly with the number of iterations and finally converge.

	 To evaluate the performance of the proposed
	schemes, we provide the numerical comparison of the WMR achieved by the following schemes: 1) \textbf{ASB allocation scheme, $\boldsymbol{\Gamma=0}$}: we solve a problem similar to problem \eqref{p1} under the assumption of perfect SI cancellation, i.e., $\Gamma=0$; 2)  \textbf{ASB allocation scheme}: problem \eqref{p1} solved by Algorithm \ref{algorithm1} proposed in Section IV;  3) \textbf{ESB allocation scheme, ub}: the upper bound in \eqref{jensen}; 4) \textbf{ESB allocation scheme,  $\boldsymbol{\Gamma=0}$}: problem \eqref{p2} solved by Algorithm \ref{alg1} proposed in Section III; 5) \textbf{ESB allocation scheme}: ESB allocation solved by the algorithm proposed in Section IV-C; 6) \textbf{Fixed sub-band bandwidth (FSB) allocation scheme}: sub-bands of equal bandwidth are  pre-allocated to users based on a predefined rule; 7) \textbf{HD ASB allocation scheme}: the HD counterpart of the scheme in 1).
	
\begin{figure}[t]
	\centering
	\includegraphics[width=2.7in]{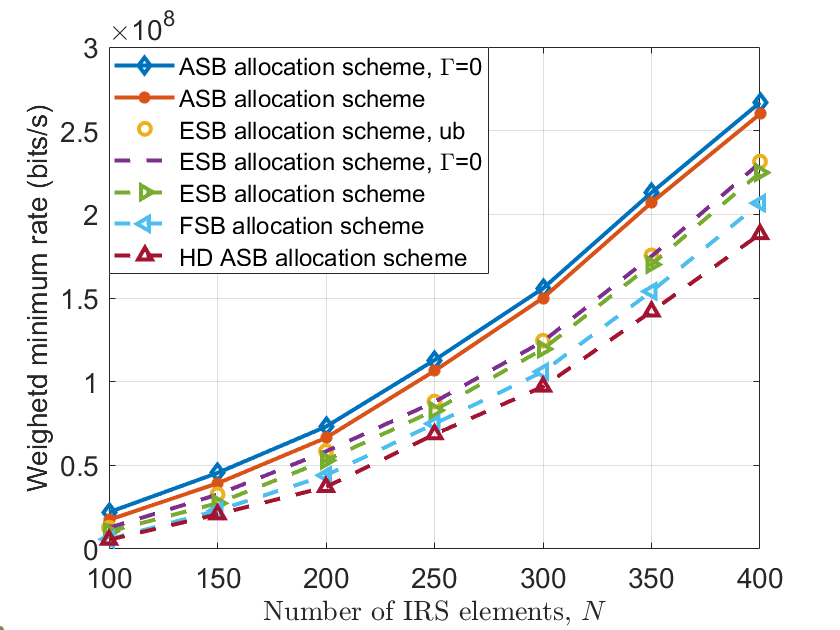}
	\vspace{-0.1cm}
	\caption{WMR versus the number of IRS elements.}
	\label{fig_N}
	\vspace{-0.3cm}
\end{figure}

	In Fig. \ref{fig_N}, we compare the WMR performance achieved by all schemes versus the number of IRS elements, $N$. As expected, the WMRs of all schemes increase with \(N\), which is attributed to the enhanced passive beamforming gain provided by a larger IRS array. By increasing \(N\), the IRS can more effectively manipulate the propagation environment, thereby improving the desired signal power at the users. Among the evaluated schemes, the proposed ASB allocation scheme consistently achieves the highest WMR across the entire range of \(N\). This performance gain becomes increasingly significant as \(N\) grows since both the ESB and FSB schemes lack joint optimization over sub-band bandwidth and resource allocation. For instance, at \(N=400\), the proposed ASB scheme achieves approximately 16\% and 34\% higher WMRs than the ESB and FSB schemes, respectively. This clearly highlights the advantage of the proposed design in fully exploiting the frequency diversity of the THz channel. Furthermore, the schemes with perfect SI cancellation, i.e., \(\Gamma=0\) show achievable WMR under ideal scenario, and the observed performance gap between such ideal case and the practical  case quantifies the impact of residual SI on system throughput. Comparing the ESB scheme with perfect SI cancellation and its upper bound, i.e., the yellow circle curve and purple dashed curve, it can be concluded that the bound in \eqref{jensen} is relatively tight, which indicates the efficiency of Algorithm \ref{alg1}.

%
%
%
%
%
%
%
%
%
%

\begin{figure}[t]
	\centering
	\includegraphics[width=2.7in]{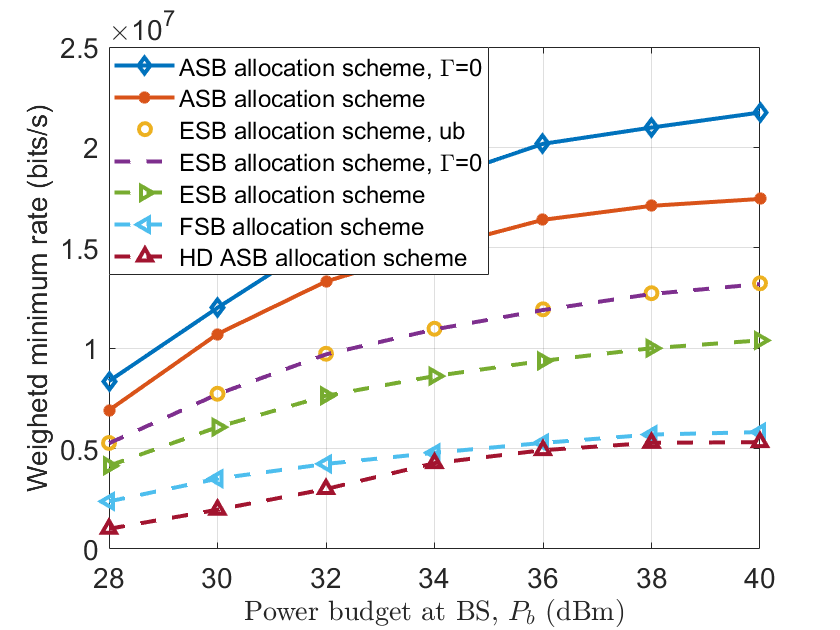}
	\vspace{-0.1cm}
	\caption{WMR versus the power budget at BS.}
	\label{fig_P}
	\vspace{-0.3cm}
\end{figure}

Fig.~\ref{fig_P} illustrates the WMR performance versus the BS transmit power budget, \(P_{\text{b}}\). As expected, the WMR of all schemes improves with increasing \(P_{\text{b}}\), which is attributed to the enhanced DL transmit power that strengthens the received signal quality at the users. The increase in \(P_{\text{b}}\) results in the growing residual SI power at the BS, which becomes a dominant factor degrading the UL reception quality at high transmit powers. Consequently, the performance gain of the perfect SI cancellation case over the non-perfect case becomes more pronounced. Another observation is that the HD ASB allocation scheme, suffering from a fundamental time-division loss, gradually catches up with the FSB allocation scheme as \(P_{\text{b}}\) increases, despite the inherent time-division loss in HD operation. This trend further highlights the superiority of efficient spectrum resource allocation in enhancing system fairness and improving the WMR, as the FSB scheme does not benefit from ASB allocation while suffers from stronger residual SI power at higher \(P_{\text{b}}\).

\begin{figure}[t]
	\centering
	\includegraphics[width=2.7in]{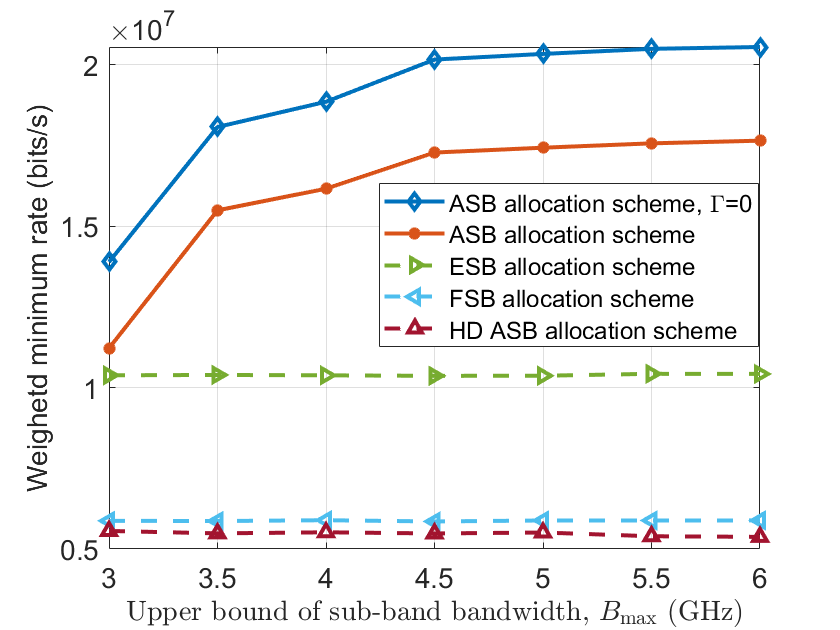}
	\vspace{-0.1cm}
	\caption{WMR versus the upper bound of sub-band bandwidth.}
	\label{fig_Bmax}
	\vspace{-0.3cm}
\end{figure}
\begin{figure}[t]
	\centering
	\includegraphics[width=2.7in]{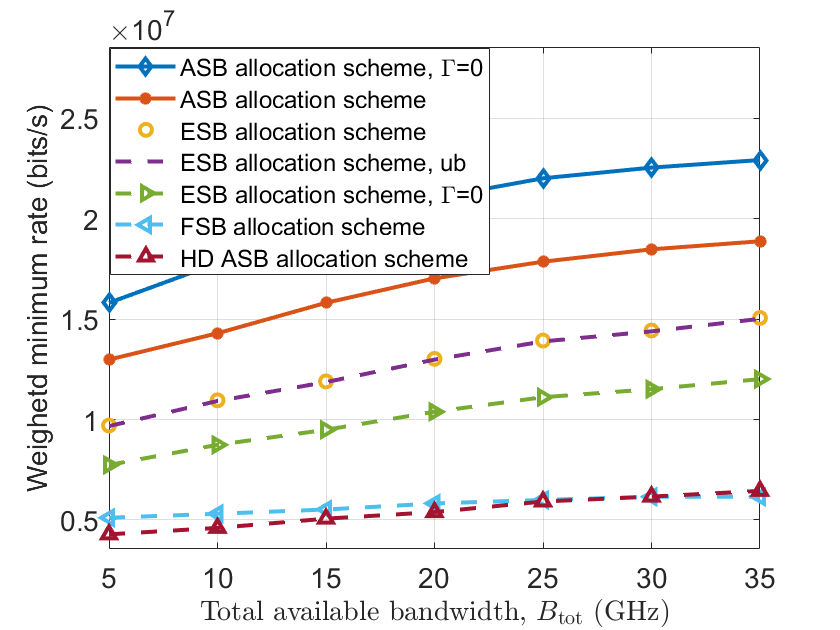}
	\vspace{-0.1cm}
	\caption{WMR versus total available bandwidth.}
	\label{fig_Btot}
	\vspace{-0.3cm}
\end{figure}
\begin{figure}[t]
	\centering
	\includegraphics[width=2.7in]{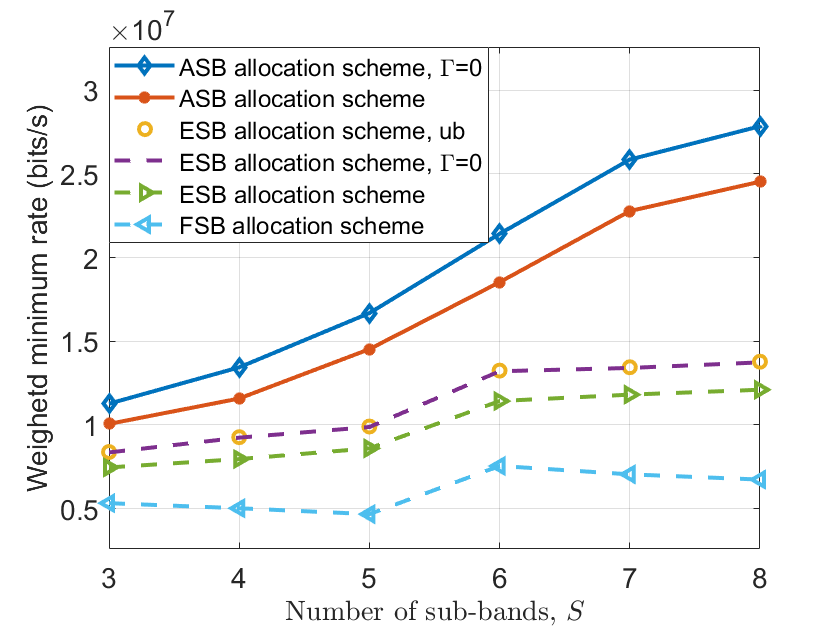}
	\vspace{-0.1cm}
	\caption{WMR versus number of sub-bands.}
	\label{fig_S}
	\vspace{-0.3cm}
\end{figure}
\begin{figure}[t]
	\centering
	\includegraphics[width=2.7in]{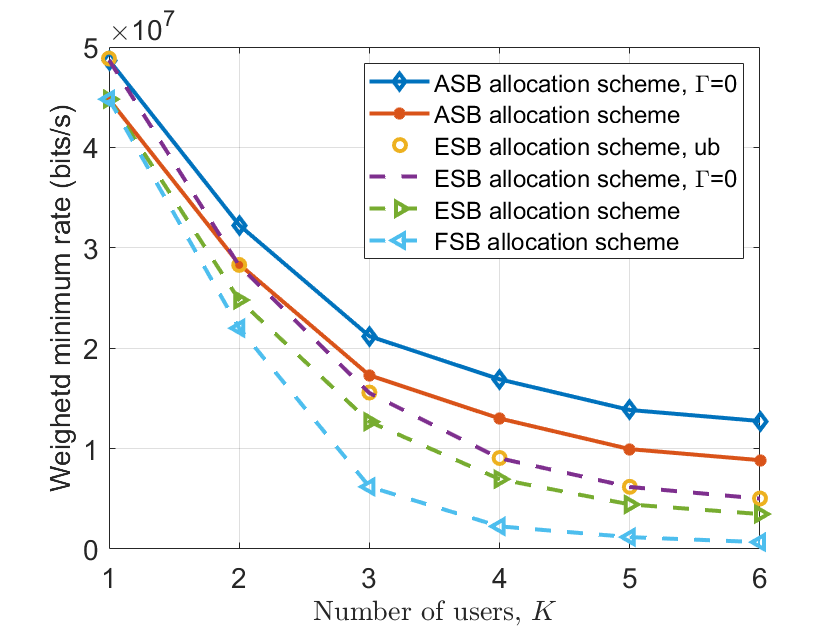}
	\vspace{-0.1cm}
	\caption{WMR versus number of users.}
	\label{fig_K}
	\vspace{-0.3cm}
\end{figure}
Fig. \ref{fig_Bmax} illustrates the WMR performance  versus the upper bound of sub-band bandwidth, $B_{\text{max}}$. As $B_{\text{max}}$ increases, both ASB allocation schemes, with perfect SIC and non-perfect SIC, demonstrate significant WMR improvements due to their greater flexibility in utilizing the available spectral resources. In contrast, the ESB and FSB allocation schemes exhibit nearly flat WMR trends, since their rigid bandwidth allocation prevents them from adapting to the increased $B_{\text{max}}$. Notably, the WMR improvement for the HD ASB scheme is much less significant compared to its FD counterparts. This is because, in the HD case, the total available spectrum must be partitioned across both UL and DL phases separately, effectively halving the spectral degree of freedom compared to FD operation. As a result, even when $B_{\text{max}}$ grows, the HD system’s limited bandwidth per phase constrains its WMR improvement.

Fig. \ref{fig_Btot} illustrates the WMR performance versus the total available bandwidth, $B_{\text{tot}}$. As observed, the WMR achieved by all schemes improves with increasing total bandwidth due to the enlarged spectral resources. Among the schemes, the proposed ASB allocation scheme consistently outperforms the other benchmarks across the entire bandwidth range, highlighting the benefit of ASB partitioning in effectively utilizing the available spectrum. Moreover, as the bandwidth increases, the performance gap between the FD ASB schemes and the other benchmarks gradually widens, emphasizing the effectiveness of joint bandwidth and beamforming optimization in wideband scenarios.
It is  noteworthy that the HD ASB scheme shows inferior performance relative to all FD-based schemes when $B_{\text{tot}}$ is not relatively small, underscoring the spectral efficiency advantage brought by FD operation. An interesting observation is that the HD ASB scheme overtakes the FSB scheme at higher $B_{\text{tot}}$. This trend can be explained by the fact that in wideband scenarios, the benefit of spectrum resource allocation becomes more dominant, allowing the HD system to better exploit the available bandwidth, thereby narrowing the gap with the less spectrum-efficient FSB scheme, which also suffers from SI. Overall, the results highlight both the importance of effective bandwidth allocation and the inherent advantage of FD operation.

Fig. \ref{fig_S} shows the WMR performance versus the number of sub-bands, $S$. The ASB allocation schemes, both with and without perfect SIC, exhibit a consistent WMR improvement as $S$ increases, benefiting from their ability to adaptively allocate varying bandwidths across the growing number of subbands. This flexibility allows them to better exploit the increased spectral resolution offered by a higher $S$. In contrast, the ESB and FSB schemes experience WMR performance degrdation when $S$ grows from 3 to 5. This is because the number of users in the system is set to 3, making 3 subbands sufficient for user separation. Moreover, both ESB and FSB schemes allocate equal bandwidth across sub-bands, so increasing $S$ without increasing the total bandwidth simply reduces the bandwidth per sub-band, resulting in performance degradation. However, a clear performance transition point appears when $S$ reaches 6, as the total bandwidth can be more evenly divided among the 3 users, leading to a more balanced resource allocation.

Fig. \ref{fig_K} depicts the WMR performance versus the number of users, $K$. As expected, the WMR of all schemes exhibits a clear downward trend with increasing $K$, due to the growing competition for limited spectral, power, and IRS resources. The ASB allocation schemes consistently outperform the ESB and FSB schemes across the entire user range. Notably, the performance gap between the ASB schemes and the benchmark schemes widens as $K$ grows. This is because the rigid bandwidth allocation in ESB and FSB cannot effectively mitigate the resource shortage caused by serving more users. Particularly, the FSB scheme suffers the most severe WMR degradation due to its complete lack of bandwidth adaptability, leading to nearly negligible WMR at $K=6$. This result highlights the scalability advantage of the proposed ASB-based designs in handling larger user loads in IRS-assisted FD THz systems.

\vspace{0.1cm}
\section{Conclusion}
In this paper, we studied an IRS-assisted FD THz communication system, where joint spatial, spectral, and power resource allocation was performed, capturing the unique FD and THz band characteristics. Specifically, a WMR maximization problem was formulated by jointly optimizing the IRS reflection beamforming, DL/UL transmit power, sub-band bandwidth allocation, and sub-band assignment, while explicitly considering the practical impact of frequency-related molecular absorption and residual SI. To balance performance and complexity, two algorithms were developed under different spectrum partitioning schemes. The first adopts ESB for simpler and tractable optimization, while the second employs ASB to dynamically allocate resources, enhancing spectral utilization and flexibility. Simulation results validated the effectiveness of both designs and further demonstrated the superiority of ASB allocation in improving spectral efficiency.

\bibliographystyle{IEEEtran}
\IEEEtriggeratref{25}
\bibliography{reference}

\end{document}